\pdfoutput=1

\documentclass[aps,pra,preprintnumbers,superscriptaddress,10pt,twocolumn,floatfix]{revtex4-2}

\usepackage[version=3]{mhchem}
\usepackage{graphicx}
\usepackage{mathtools}
\usepackage{mathrsfs}
\usepackage{txfonts}
\usepackage{upgreek}
\usepackage[colorlinks=true,citecolor=blue,urlcolor=red,verbose]{hyperref}
\usepackage{lipsum}
\usepackage{tensor}
\usepackage[utf8]{inputenc}
\usepackage[T1]{fontenc}
\usepackage[exponent-product = \cdot]{siunitx}
\usepackage{cleveref}           
\usepackage{natbib}
\usepackage{xcolor}
\usepackage{array,multirow,graphicx}
\usepackage{booktabs}
\usepackage{color}
\usepackage[english]{babel}

\begin{document}
\sisetup{tight-spacing=true}
\setcitestyle{super}

\allowdisplaybreaks

\title{Statics and Dynamics of Space-Charge-Layers in Polarized Inorganic Solid Electrolytes} 

\author{Katharina Becker-Steinberger}
\affiliation{Institute of Engineering Thermodynamics, German Aerospace Center (DLR), Pfaffenwaldring 38-40, 70569 Stuttgart, Germany. }
\affiliation{Helmholtz Institute Ulm (HIU), Helmholtzstra{\ss}e 11, 89081 Ulm, Germany.}
\email{katharina.becker-steinberger@dlr.de}
\author{Simon Schardt}
\affiliation{University of W\"{u}rzburg, Campus Hubland Nord 32, 97074 W\"{u}rzburg, Germany.}
\author{Birger Horstmann}
\affiliation{Institute  of  Engineering  Thermodynamics, German Aerospace Center (DLR), Pfaffenwaldring 38-40, 70569 Stuttgart, Germany. }
\affiliation{Helmholtz Institute Ulm (HIU), Helmholtzstra{\ss}e 11, 89081 Ulm, Germany.}
\affiliation{Ulm University, Albert-Einstein-Allee 47, 89081 Ulm, Germany.}
\author{Arnulf Latz}
\affiliation{Institute  of  Engineering  Thermodynamics, German Aerospace Center (DLR), Pfaffenwaldring 38-40, 70569 Stuttgart, Germany. }
\affiliation{Helmholtz Institute Ulm (HIU), Helmholtzstra{\ss}e 11, 89081 Ulm, Germany.}
\affiliation{Ulm University, Albert-Einstein-Allee 47, 89081 Ulm, Germany.}

    \begin{abstract}
      The quest for safe high-energy batteries with ``5V-class'' cathodes and lithium metal anodes drives research into solid electrolytes. However, reasons for the large charge transfer resistances---the major bottleneck of all-solid-state batteries---are still debated. 
In this article, we explore the processes in incompressible solid electrolytes between blocking electrodes by theory-based continuum modeling and numerical simulations. We investigate the experimentally observed wide space-charge-zones in solid electrolytes, which are a possible cause for the high interfacial resistances. 
On time scales relevant for battery applications, we reduce our model equations. 
Analytic and numeric calculations predict and study the actual structure of space-charge-layers in solid electrolytes. To illustrate these dynamics and validate our model, computational results are presented and compared with experimental observations. 
Analog to semiconductors, we determine the material dependent, asymmetric space-charge-layer width in the low temperature limit approximately. 
This allows us to make an explicit statement about the influence of defect concentrations and dielectric properties on the width of the space-charge-layers in homogeneous solid electrolytes.
    \end{abstract}
  \maketitle

\section{Introduction}
All-solid-state batteries (ASSBs) are experiencing a growing scientific interest from both practical and theoretical
points of view in recent years as potential 
next-generation high-voltage batteries with great intrinsic safety features. 
Many of the limitations and risks of current technologies can be addressed by using an inorganic solid electrolyte (SE) in place of the standard, flammable organic liquid electrolytes. Since SEs are intrinsically non-flammmable, there is no risk of thermal runaway and leakage. 
Apart from that, SEs appear to be chemically more stable and show wide voltage windows~\cite{Dudney_2005, Kamaya_2011, Liu_2013, Sahu_2014,Rangasamy_2014}.
Some SEs are even compatible with lithium metal anodes. 
Both, enhanced safety and higher voltages make ASSBs an attractive candidate for automotive applications~\cite{BMW, Toyota}.

However, ASSB cells still face essential difficulties. For a long time, the major drawback of single-ion conducting inorganic solids was their low ion conductivity compared to organic liquid electrolytes. Recent developments of super-ion conductors overcame these drawbacks and exhibit ion conductivities even higher than those of liquid electrolytes~\cite{Kamaya_2011}.  Instead, low interfacial charge transfer rates and low power densities~\cite{Luntz_2015, Zhu_2016, Richards_2016} still pose a problem and the 
reasons for the large charge transfer resistances are still subject to scientific discussions.
Different mechanisms are discussed. Among them are thermodynamic instabilities at the SE-electrode interfaces causing different types of interfaces and extended interphases~\cite{Wenzel_2016}, mechanical issues and small effective contact areas~\cite{Luo_2017} for charge transfer reactions as well as lithium-ion depletion and accumulation zones at the interface~\cite{Maier_1995}, usually referred to as space-charge-layers (SCLs).  
Most probably there is no unique answer to the question of the cause of interfacial resistances for all material systems. 
A fundamental and more detailed understanding of the phenomena and their interplay is required as a prerequisite for the discussion of their relevance for interfacial resistance in ASSBs.

SCLs in SEs have the same origin as double layers in liquid electrolytes.  
But in ASSBs with inorganic SEs, both the formation process and the resulting SCLs differ significantly from liquid electrolytes due to the fixed anion lattice structure.
\emph{In situ} electron holography measurements of the electric potential distribution around the interfaces during polarization experiments indicate 
that the potential drop at the interface is much wider than in liquid electrolytes~\cite{Yamamoto2010, Hirayama_2017, Aizawa_2017, Nomura_2019}. Possible causes for extended SCLs were presented in Ref.~\cite{Braun_2015}.
However, the width and the role of SCLs in ASSBs is controversially discussed~\cite{deKlerk_2018, Brogioli_2019}.
In this discussion qualitative and quantitative dependencies and statements are mixed. 

Free energy based theories using rigorous techniques from non-equilibrium thermodynamics to derive the mathematical models for equilibrium and non-equilibrium situations are necessary to create an unbiased picture of the expected physical processes. 

Computational modeling and theoretical investigations on multiple scales are important tools to understand the different processes and mechanisms. Examples for recent works range from atomic scale\cite{Zhu_2016, Miara_2016, Yu_2017,Fingerle_2017, Swift2019} to continuum models and theories\cite{Braun_2015, Maier_2017, Monroe_2017,Tian_2017, Srinivasan_2017, Finsterbusch_2018, deKlerk_2018, Brogioli_2019, Li_2019, Neumann_2020, Liu_2020, Li_2020, Hao_2020}. The continuum approaches can be methodically divided into (i) phenomenological models\cite{Maier_2017,  deKlerk_2018, Brogioli_2019,Liu_2020, Hao_2020}, (ii) free energy based models\cite{Braun_2015}, and (iii) mixed forms\cite{Li_2019, Li_2020}. On the other hand, depending on the spatial resolution at the interfaces, one can distinguish between near-interface-resolved models\cite{Braun_2015, Maier_2017, deKlerk_2018, Li_2019, Liu_2020} and bulk-type approaches\cite{Finsterbusch_2018,Neumann_2020}. Bulk-type models allow the simulation of micro-structured composite electrodes and the study of cell architectures and designs. Near-interface-resolved approaches are particularly well suited for investigating the coupling of local charge imbalances, electric fields and electromechanical forces in the SCLs and their effect on the charge transfer reaction at the interface.

It is also common to set the near interface resolved approaches, in relation to the Poisson-Nernst-Planck (PNP) theory. The original PNP theory~\cite{Nernst_1889, Planck_1890a, Planck_1890b} is based on a simple physical picture of non-interacting diffusing ions obeying Boltzmann statistics and the Poisson equation to account for the charges of the ion. The corresponding classical equilibrium theory~\cite{Debye_1928, Chapman_1913}, describes non-interacting particles by a Poisson-Boltzmann equation. The width of the SCLs predicted by this theory is characterized by the Debye length. Transport models that deviate from the classical PNP theory are often called modified PNP models, regardless of the method used to derive these models. This distinction is primarily historical and is due to the widespread use of PNP approaches.

In the late 1970s Kornychev and Vorotyntsev~\cite{Vorotyntsev_1976, Kornychev_1981} introduced a modified PNP theory for transport and SCLs in homogeneous SEs by considering the effect of finite lattice sites. In contrast to the classical PNP approach, Kornychev and Vorotyntsev~\cite{Vorotyntsev_1976, Kornychev_1981} derived a Fermi-Dirac like distribution of cations in SEs in analogy to semiconductors. In the following years, both standard approaches and similar phenomenological modifications of the approaches, have been used to study ASSB cells or SEs~\cite{Maier_1995, Becker_2010, Landstorfer_2011, Danilov_2011}. These  
approaches have also been combined with phenomenological concepts on the structure of SCLs developed for liquid electrolytes, such as extending the diffuse SCL by a compacted Stern layer of fixed thickness~\cite{Brogioli_2019, Becker_2010, Landstorfer_2011}. 

Using a free energy based theory for a liquid electrolyte, Dreyer et al.~\cite{Dreyer_2013} showed that PNP cannot be applied for larger ion concentrations and that coupling of momentum
and mass conservation equations leads to a deviation from Boltzmann statistics. Consequently Nernst-Planck-based approaches to ion transport fail at the interfaces because they neglect the high pressures induced by Maxwell stress and ignore species interactions and volume constraints. Nevertheless, PNP approaches and their phenomenological modifications for the investigation of liquid and solid electrolytes are still very common and, for example, used to study SCLs induced by grain boundaries\cite{GoebelI_2014, GoebelII_2014, Maier_2017} and Coulomb interactions in solids~\cite{deKlerk_2018}.

\begin{figure}[!tb]
\begin{center}
\includegraphics[width=0.3\textwidth]{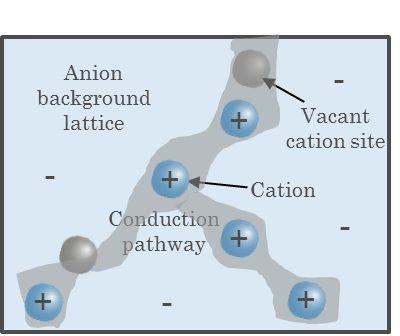}
\caption{Schematic illustration of inorganic SEs structure consisting of a fixed anion background lattice (blue area) and mobile cations (light blue dots). The long-term transport of cations follows conduction pathways (gray area) via empty cation sites (gray dots).}
\label{Fig_SchematicPathways}
\end{center}
\end{figure}
In a previous work, Braun et al.~\cite{Braun_2015} developed a free energy based model for SCL formation in incompressible, dielectric, monocrystalline SEs between two ion-blocking electrodes. Their model predicts a cation site limited, asymmetric equilibrium cation distribution in the SCLs, which is qualitatively similar to the one found by Kornychev and Vorotyntsev~\cite{Vorotyntsev_1976, Kornychev_1981}. The width of the equilibrated SCLs was found to be much larger than the Debye length. 

In this article, we (i) apply the model derived in Braun et al.~\cite{Braun_2015} to homogeneous and nonhomogeneous ionic crystals, (ii) extend it by considering a general coupling between mechanics and configuration, (iii) derive analytical expressions characterizing SCLs in equilibrium, and (iv) study dynamical processes for SCLs on time scales relevant for battery applications. 
\begin{itemize}
\item The general coupling between mechanics and configuration allows  considering different ion sizes and lattice distances. 
For spatially homogeneous SEs without ion size effects we obtain an equilibrium cation distribution in the SCLs qualitatively similar to the one postulated by Kornychev and Vorotyntsev~\cite{Vorotyntsev_1976, Kornychev_1981}. Any deviation from homogeneity and uniformity leads to fundamentally different distributions.
Our approach thus enables new physical interpretations of the earlier model~\cite{Vorotyntsev_1976, Kornychev_1981} by embedding it in a more general theoretical frame. 
\item By means of the analytical solution for the bulk potential value we now can analytically identify the potential drop across the SCLs. Furthermore, we find explicit expressions for the width of SCLs in SEs as function of material parameters. Our analytical solutions thus allow us to rationalize the qualitative results found earlier~\cite{Braun_2015} and provide insight into the possible origin of experimentally observed large screening lengths in SEs. 
\item We complete the picture with dynamic simulations.
\end{itemize}
This work is organized as follows. First, the modeling is described in detail in Sec.~\ref{Sec_Theory}. In Sec.~\ref{Sec_SimulationDetails}, we briefly comment on the materials investigated. 
Then, we present in Sec.~\ref{Sec_EquilibriumResults} our theoretical and simulations results on SCLs in polarized SEs. 
The paper ends with a conclusion in Sec.~\ref{Sec_Conclusion}. 
Further background information on the modeling, computational preliminaries, and analytical results are provided in the supplementary information.   

\begin{figure}
\begin{center}
\includegraphics[width=0.3\textwidth]{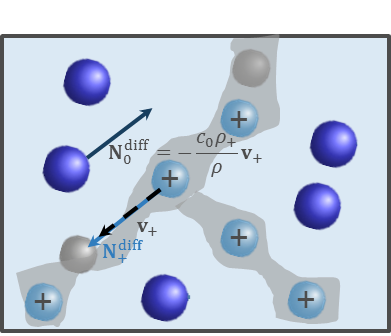}
\caption{Schematic illustration of diffusion with respect to the center-of-mass motion through cation motion.  Cations (light blue dots) move at a velocity $\textbf{v}_{+}$ (dashed black arrow). The background lattice anions (blue dots)  is  immobile. Mass is transported as convection of the cation masses  and mass diffusion of cations (light blue arrow).  Both mechanisms are proportional to each other.  The mass diffusion flux of anions (dark blue arrow) is compensated by their mass convective motion.}
\label{Fig_CenterOfMassMotion}
\end{center}
\end{figure}

\section{Mathematical Model}
\label{Sec_Theory}
In this section, we derive a model for cation conducting incompressible, isotropic elastic, dielectric inorganic SEs between two ion-blocking electrodes using fundamental principles from non-equilibrium thermodynamics. It describes the thermo-electrodynamic response of incompressible SEs to polarization experiments\cite{Hirayama_2017, Aizawa_2017, Nomura_2018} as function of time and space on time scales relevant for most battery applications. 

As a preliminary remark, Sec.~\ref{Sec_Assumption} outlines the types of assumptions that the model derivation includes.
Starting point of our actual derivation, in Sec.~\ref{Subsec_SEModel}, is a thermodynamically consistent incompressible SE model based on previous work~\cite{Braun_2015}. We consider this model for both spatially homogeneous and heterogeneous materials. In Sec.~\ref{Subsec_dDDP} we show by a time scale analysis of processes that the system of equations can be reduced. The different incompressible scenarios result in qualitatively different driving forces. 
Then, we state the initial and boundary conditions for the polarization experiments in Sec.~\ref{Subsec_IC}. Finally, we formulate the equations characterizing thermodynamic equilibrium in Sec.~\ref{Subsec_EQ}.

\subsection{Model Assumptions}
\label{Sec_Assumption}
The derivation of the model contains assumptions regarding setup, domain geometry, interfaces, material composition and modeling, as well as processes and time scales. A summary of these assumptions can be found in the supplementary information (see Sec.~SI-1). The assumption of blocking electrodes allows us to study the multi-physical process of SCL formation due to polarization without superimposing the process with others such as interface kinetics.

\subsection{Modeling Incompressible Solid Electrolytes}
\label{Subsec_SEModel}
In the following, we derive the different model building blocks for self-consistent cation and charge transport in electric fields in inorganic incompressible SEs using non-equilibrium thermodynamics. 
Special emphasize is put on the implications of a general coupling between mechanical and configurational contributions to the free energy on the structure of the cation diffusion flux. 
All notations are given in Tab.~\ref{Tab_Constants} and Tab.~\ref{Tab_Symbols}.

\subsubsection{Constituent Model for Fast Single Ion Conductors}
Ion transport in solids depends on the lattice structure. On an atomic scale ion diffusion in solids is given by a series of activated jumps in the lattice structure.
For detailed atomic transport mechanisms and background on solid ion conductors, we refer to the textbooks of Mehrer\cite{Mehrer_2007} and Maier\cite{Maier_2004}.
The long-term transport in many fast
single ion conductors follows connected ionic conduction pathways, schematically illustrated in Fig.~\ref{Fig_SchematicPathways}. 
They may be considered as the connected set of energetically available sites for the ion. In principle, these pathways can be identified and calculated on an atomic scale by both first-principle\cite{Li_2010, Geiger_2011, Takahashi_2018} and bond valence methods\cite{Thangadurai_2004, Adams_2012, Ehrenberg_2016, Stoeffler_2018}.
For describing transport on the macroscopic scale only the thermodynamics and kinetics of these accessible sites is relevant for developing a proper transport theory. More specific, we model these SEs as three component systems ($N=3$) consisting of mobile cations (indexed +), vacant cation sites neglecting possible charges (indexed $v$), and a negatively charged fixed and incompressible background skeleton (indexed $0$), which is formed by immobile anions. We note that this model picture applies to inorganic SEs with intrinsic defects as well as amorphous glasses. 

The individual properties of the constituents are characterized by molar masses $M_{\alpha}$,  charge numbers $z_{\alpha}$, molar densities (referred to as concentrations) $c_{\alpha}$, velocities measured in a lab frame $\textbf{v}_{\alpha}^{\#}$, and molar volumes $\nu_{\alpha}$. In contrast to the atomic constituents, the cation sites are mass-less and charge-less, i.e., $M_v=0, z_v=0$. Since the anions are resting in the background lattice and lattice deformations are neglected (see assumption (SI-M1) and (SI-M2)), the anions do not have a velocity, i.e., $\textbf{v}_0^{\#}=0$, and the fixed,
charged background concentration $c_0$ is a time invariant. The polarization of the electrolyte (see assumption (SI-S1)) causes only the movement of the cations and the general cations sites each at a velocity $\textbf{v}_{\alpha}^{\#}, \alpha \in\{+,v\}$.

\subsubsection{Reference Frame}
Transport processes can be described in different reference frames. Since we focus on the importance of momentum conservation for deriving proper transport theories in systems with space charges, we chose as in Braun et al.~\cite{Braun_2015} the  center-of-mass velocity as independent variable. Due to the fixed anion lattice, any motion of the cation leads naturally to a center-of-mass motion i.e., to convection. 
Due to the assumed incompressibility of the background lattice, the internal center-of-mass velocity, which mass averages the individual constituent velocities, becomes~\cite{Braun_2015}
\begin{equation}
\textbf{v} = \frac{\rho_+ \textbf{v}_+^{\#} + \rho_0 \textbf{v}_0^{\#}}{\rho} = \frac{\rho_+}{\rho} \textbf{v}_+^{\#}, \label{v_com}
\end{equation}
where $\rho_{\alpha} =M_{\alpha}c_{\alpha}, \alpha \in \{+,0\},$ are the constituent mass densities and $\rho=\sum_{\alpha\in \{+,0\}}\rho_{\alpha}$ is the total mass density. This center-of-mass velocity is determined by the momentum equation~\cite{degroot_1969}.

For consistency, we study the different contributions to the cation and cation site transport in a center-of-mass frame. A formulation in lattice-fixed coordinates is given in the supplementary information, see Sec. SI-1.2.\footnote{Throughout this work, quantities or parameters associated with the lattice-fixed frame are marked with superscript $\#$.}
Therefore, the total fluxes $\textbf{N}_{\alpha}^{\#} = c_{\alpha}\textbf{v}_{\alpha}^{\#}, \alpha\in\{+,v\},$ are separated into convective fluxes $c_{\alpha}\textbf{v}$ and non-convective fluxes $\textbf{N}_{\alpha}=c_{\alpha}(\textbf{v}_{\alpha}^{\#}-\textbf{v})$. The convection describes the motion of the center-of-mass. As a consequence of this perspective, we have to consider two ionic diffusion fluxes of cations and anions relative to the convective center-of-mass motion, respectively. This is illustrated in Fig.~\ref{Fig_CenterOfMassMotion}. Since the anions are resting in the lab frame, the center-of-mass velocity (\ref{v_com}), the cation velocity (dashed black arrow) and diffusion flux (blue arrow) are proportional to each other\cite{Braun_2015}
\begin{align}
\textbf{N}_+ =c_{+}\left(\textbf{v}_+^{\#} - \textbf{v}\right) =  c_+\frac{\rho_0}{\rho} \textbf{v}_+^{\#} =  c_+\frac{\rho_0}{\rho_+} \textbf{v}. \label{FluxCorrelation}
\end{align}
Due to mass conservation, the mass diffusion fluxes $M_{\alpha}\textbf{N}_{\alpha}$ sum up to zero and  
\begin{align}
\textbf{N}_{0}= -m_+\textbf{N}_{+} \label{MassConservationSideCondition}
\end{align}
is a dependent quantity, where $m_+:=M_+/M_0$ abbreviates the ratio of molar masses.

\begin{figure}
\begin{center}
\includegraphics[width=0.4\textwidth]{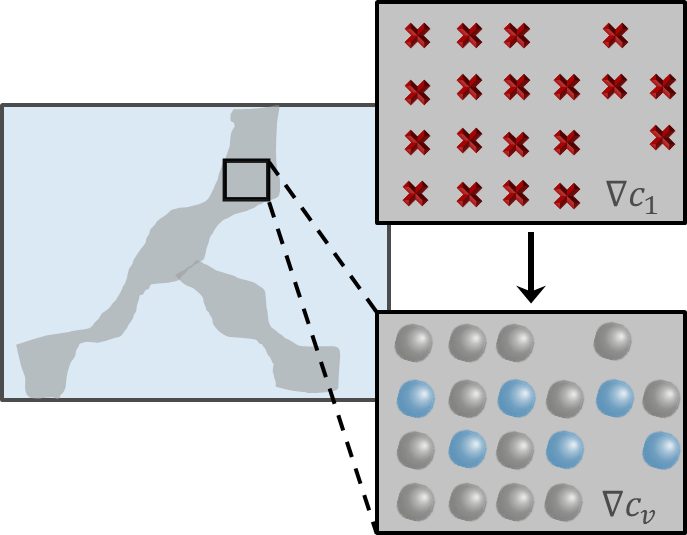}
\caption{Sketch of the influence of possible spatial inhomogeneities on the cation sites. Due to material production process the overall cation sites (red crosses) may spatially variate (top left). On these cation sites the same amount of cations (blue dots) are distributed resulting in local inhomogeneities on the conduction pathways.}
\label{Fig_Inhom}
\end{center}
\end{figure}

\subsubsection{Transport Equations}
In this paper, we consider incompressible SEs with possible local inhomogeneities (see Eq. (\ref{DeltaNu})) caused by the material manufacturing process or the cell assembly. Therefore, the concentration of accessible cation sites on the conduction pathway $c_1=c_++c_v$ may vary locally. As shown in Fig.~\ref{Fig_Inhom}, these local material variations are associated with local defect concentration gradients. 
As a further consequence of incompressibility, the vacancy concentration and the mass density evolution are completely determined by the cation evolution. This cation evolution obeys the continuity equation
\begin{align}
\partial_t c_{+} +\nabla\cdot\left(c_{+} \textbf{v}_{+}^{\#}\right) &=0.
\end{align}

As a consequence the equation for cation transport is reduced to a generalized diffusion equation for the cations and a momentum balance\cite{Braun_2015} 
\begin{alignat}{3}
\partial_tc_+ +\nabla \cdot \left(\frac{\rho}{\rho_0}\textbf{N}_+\right)&=0,\label{Braun1a}\\
\partial_t\rho\textbf{v} +\nabla\cdot(\rho\textbf{v}\otimes\textbf{v}- {\cal T})&=0,\label{Braun2a}
\end{alignat}
where ${\cal T}$ denotes the stress tensor.

In addition to these ion transport equations, we need equations for charge balances, i.e., equations for electric charge evolution and electric currents, and their coupling to the electromagnetic fields.
Neglecting magnetic effects  (see assumption (SI-P2)) Maxwell's equations reduce to Coulomb's equation 
\begin{align}
\nabla \cdot \textbf{D} &= q_F, \label{Coulomb}
\end{align}
where $q_F=F\sum_{\alpha \in\{+,0\}}z_{\alpha}c_{\alpha}$ is the free charge density and $F$ the Faraday constant. As usual, the dielectric displacement may be written in terms of the polarization $\textbf{P}$,  i.e., $\textbf{D} = \epsilon_0\textbf{E}+\textbf{P}$ with $\epsilon_0$ being the vacuum dielectric constant. The electric field can be expressed as gradient of a scalar potential, only, $\textbf{E}=-\nabla\phi$. 
\begin{figure*}
\begin{center}
\includegraphics[width=0.6\textwidth]{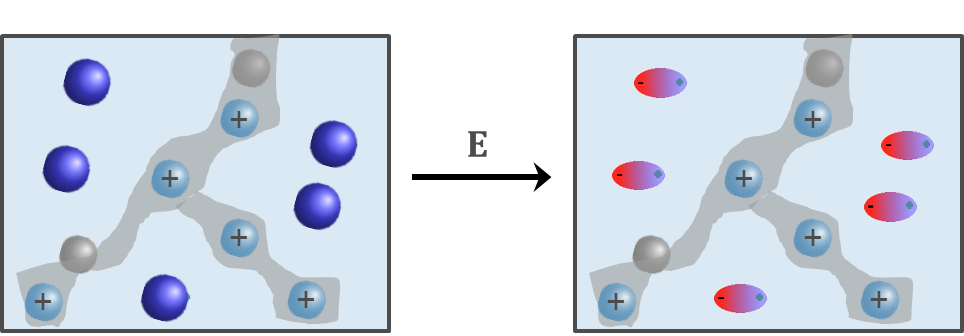}
\caption{Schematic description of the polarization current in a monocrystal. The anion lattice (blue dots) is polarized by an applied electric field (dipoles with color gradient).}
\label{Fig_Polarization}
\end{center}
\end{figure*}

The total electrical charge density $q$ consists of this free charge density and an additional polarization density~\cite{Truesdell}. Accordingly, the measurable or experimentally controlled electric current $\textbf{j}^{\#}$ is also generated by free charge currents and local polarization changes. 
We assume that only the background lattice is polarizable (see assumption (SI-M3)). Fig.~\ref{Fig_Polarization} shows a schematic of this assumption. The presence of an external electric field induces a polarization of the anion lattice by local shifts of bound charges (dipoles with color gradient). These time changes of the polarization give rise to a polarization current $\textbf{j}_p^{\#}=\partial_t\textbf{P}$. 
On the contrary, the free charge current density  
is only caused by the mobile cations, i.e., $\textbf{j}_F^{\#}=z_+Fc_{+}\textbf{v}_+$. Charge conservation implies 
\begin{align}
 \partial_t q +\nabla\cdot\textbf{j}^{\#}=0.\label{ChargeConservation}
\end{align}
Similar to the total species fluxes, we split the charge transport contributions in Eq. (\ref{ChargeConservation}) relative to the internal center-of-mass velocity. Under assumption (SI-M3) there is no flux of polarization density. But due to relation (\ref{FluxCorrelation}) the free charge current density contains the same inverse weighting factor as the cation conservation law (\ref{Braun1a}). The total electric current becomes 
\begin{align}
\textbf{j}^{\#}= z_{+}F\frac{\rho}{\rho_0}\textbf{N}_{+} + 
\partial_t \textbf{P}.\label{PolarizationCurrent}
\end{align}
\subsubsection{General Structure of Constitutive Laws} 
The polarization $\textbf{P}$, the molar diffusion fluxes $\textbf{N}_{\alpha}^{\text{diff}}$, and the stress tensor ${\cal T}$, are material dependent quantities, derived from a material model and entropy principles. This material model is given by a general free energy functional of the species concentrations and the electric
field $\rho \psi =\rho \psi ((c_{\alpha})_{\alpha\in \{+,v,0\}}, \textbf{E})$. Evaluating the entropy principle in a polarizable system\cite{Dreyer_2013,Braun_2015} results in three main guide lines for thermodynamic consistency: Firstly, it leads to the equilibrium constraints $\textbf{P} =-\partial \rho \psi/\partial \textbf{E}$, $\mu_{\alpha}:=\partial \rho \psi/\partial c_{\alpha}$, which relate the polarization and the molar chemical potential $\mu_{\alpha}$ to their conjugate constitutive variables and serve as constitutive relations. 
Secondly, it gives the representation of the stress tensor as the sum of the mechanical stress and the Maxwell stress due to the electric field. 
Thirdly, for isothermal processes (see assumption (SI-P1)), the entropy principle gives rise to a dynamic constraint---the entropy inequality---for the independent thermodynamic fluxes and diving forces
\begin{align}
\zeta =-\frac{1}{\theta}\sum_{\alpha\in \{\pm, v\}}(\nabla\mu_{\alpha}-z_{\alpha}F\textbf{E})\cdot \textbf{N}_{\alpha}\ge 0,
\label{EntropyInequality1}
\end{align}
where $\zeta$ is the entropy production and
$\theta$ the constant temperature.
Using the mass conservation constraint (\ref{MassConservationSideCondition}) results in 
$N-1=2$ independent fluxes and driving forces relative to the center of mass velocity~\cite{degroot_1969} 
\begin{align}
\zeta =\frac{1}{\theta}\left(\left(\textbf{X}_{+} -m_+\textbf{X}_{0} \right)\cdot  \textbf{N}_{+} + \textbf{X}_{v}\cdot  \textbf{N}_{v}\right) \ge 0,
\label{EntropyInequality2}
\end{align}
where $\textbf{X}_{\alpha}= -(\nabla\mu_{\alpha}+z_{\alpha}F\nabla\phi), \alpha \in\{+,0\},$ are the negative electrochemical potential gradients of ions and $\textbf{X}_v=-\nabla \mu_v$ the negative chemical potential gradients of the uncharged vacancies. In other words, under the assumption made in this work, dissipation is only caused by diffusion of cations and vacancies relative to center-of-mass motion.  

To guarantee non-negativity of the entropy production, the two independent diffusion fluxes $\textbf{N}_+$ and $\textbf{N}_v$ have to be chosen such, that the expression 
(\ref{EntropyInequality2}) is a positive semi-definite binary form. 
This is assured by a linear Onsager ansatz~\cite{Onsager_1931}, which defines them as a linear combination of the driving forces
\begin{align}
\textbf{N}_{\alpha} ={\cal L}_{\alpha+} (\textbf{X}_{+}-m_+\textbf{X}_{0}) + {\cal L}_{\alpha v}\textbf{X}_{v},\qquad \alpha\in\{+,v\}.\label{LinearCombinations}
\end{align}
The $N(N-1)/2 = 3$ independent coefficients ${\cal L}_{\alpha\beta}$ of these linear combinations are the components
of the symmetric, semi-definite Onsager matrix relative to the center of mass motion.

\subsubsection{Duality Condition and Flux-Force Relation}
In incompressible solids, each jump of an ion corresponds to a vacancy moving in the opposite direction; consequently, in a three-component system with one lattice fixed component, there is only one driving force and one flux. However, the form of this driving force is not a priori known. To the best of our knowledge this was first shown in Braun et al.~\cite{Braun_2015} by deriving an additional side condition---a relation between the cation and the defect diffusion flux---from the time independence of $c_1$ (for details see Sec.~SI-1.2.3).
This duality condition 
  \begin{align}
  \textbf{N}_v  =- \beta(c_1)\textbf{N}_+,\label{SideConditions}
 \end{align} 
 where $\beta(c_1) :=1+ m_+ c_1/c_0$ abbreviates the composition dependent factor,
 complies with the statement 
 \begin{align}
 \textbf{v}_v^{\#} =  \frac{c_+}{c_v}  \textbf{v}_+^{\#} 
 \end{align}
  about the relation between the cation velocity and cation site velocities and reduces one degree of freedom. Inserting the duality condition (\ref{SideConditions}) in the entropy  inequality (\ref{EntropyInequality2}) gives that  there is only one independent center of mass diffusion flux density $\textbf{N}_{+}$ 
and one effective driving force 
\begin{align}
\begin{split}
\overline{\textbf{X}}:=&\,\textbf{X}_{+}-m_+\textbf{X}_{0}-\beta(c_1)\textbf{X}_{v}\\ 
=&\,-\left(\nabla\mu_+ -m_+\nabla\mu_0-\beta(c_1)\nabla\mu_v+ \overline{z}F\nabla\phi\right), \label{DrivingForce}
\end{split}
\end{align}
where $\overline{z} :=z_+ -m_+z_0$ abbreviates the effective charge number for the long-term transport mechanism in solids. Note that due to the duality of cations and cation sites (\ref{SideConditions}), the effective driving force (\ref{DrivingForce}) relative to the center of mass velocity depends not only on the electrochemical potential gradients of the cations and of the fixed anion background lattice but also on the chemical potential inhomogeneities of the cation sites. 

Moreover, combining the flux-force relations (\ref{LinearCombinations}) and the duality condition (\ref{SideConditions}) implies that both the cross-species mobilities ${\cal L}_{+v}={\cal L}_{v+}$  and the self-mobility of vacancies ${\cal L}_{vv}$ are related to the cation self-mobility ${\cal L}_{++}> 0$ and there is only one independent kinetic coefficient
\begin{subequations}\label{Onsager}
\begin{align}
{\cal L}_{vv}&={\cal L}_{++}\beta^2,\label{Onsager2}\\
{\cal L}_{+v}&={\cal L}_{v+} =-{\cal L}_{++}\beta. \label{Onsager1}
\end{align}
\end{subequations}
Because of the mass effects contained in the duality factor it holds $\beta(c_1)> 1$. Thus, it follows from Eq. (\ref{Onsager2}) that vacancies among each other are more mobile than cations among each other. However, relation (\ref{Onsager1}) means that both the cations flux and the cation site flux are reduced by superposition effects and the interaction between cations and cation sites. This way, we arrive at the cation flux-force relation
\begin{align} 
\textbf{N}_+ ={\cal L}_{++}\overline{\textbf{X}}.
\label{CationFlux1}
\end{align}

\subsubsection{Structure of Driving Force and Transport Coefficients}
Before we formulate a specific material model, we analyze the implications of a general coupling between mechanical and configurational contributions to the free energy on the flux-force relation (\ref{CationFlux1})
qualitatively. Here we are mainly interested in the driving force contributions and the form of the transport parameters induced by this coupling in isotropic elastic SEs with possible structural inhomogeneities. Therefore, we assume that  
the chemical potentials of the mobile species are 
general functions of the composition, the possible material inhomogeneity, and the elastic pressure $p$, whereas the chemical potential of the fixed background skeleton is a function of the elastic pressure only, i.e.,  
\begin{align}
\mu_{\alpha} =\mu_{\alpha}(c_{\alpha}, c_1, p), \, \alpha\in\{+,v\}\qquad\text{and}\qquad\mu_{0} =\mu_0(p). \label{GeneralChemicalPotentials}
\end{align}

As usual, we apply chain rule to separate different contributions from the chemical potential gradients to the driving force. Furthermore, we utilize that by incompressibility composition changes can be traced back to the cation concentration $c_+$ and the material property $c_1$.
Considering inhomogeneous SEs 
this yields for the mobile species 
\begin{align}
\nabla \mu_{\alpha} &=\frac{\partial \mu_{\alpha}}{\partial c_{+}}\bigg|_{p=p^{\circ}}\nabla c_{+} + \frac{\partial \mu_{\alpha}}{\partial c_{1}}\bigg|_{p=p^{\circ}}\nabla c_{1}  + \frac{\partial \mu_{\alpha}}{\partial p}\bigg|_{c=c^{\circ}}\nabla p, \label{chemicalPotentialGradientGeneral}
\end{align}
where $c^{\circ}$ denotes the reference summary concentration at reference pressure $p^{\circ}$. Thus, local gradients in chemical potentials of cations and  cation sites may be caused by gradients of the cation concentration, the elastic pressure, or by the material inhomogeneity, expressed as spatial gradient in the sum of occupied and unoccupied cation sites, $c_1$. The latter is a constant in time and therefore induces an inhomogeneous equilibrium distribution of cations and cation sites.  

Substitution of Eq. (\ref{chemicalPotentialGradientGeneral}) into the cation flux driving force (\ref{DrivingForce}) allows us to 
identify canonical expressions for the transport coefficients of the processes induced by these gradients: The chemical diffusion coefficient $D_+$ is the linear combination of kinetic coefficients and thermodynamic coefficients. 
Using the thermodynamic correlation 
\begin{align}
\nu_{\alpha} =\frac{1}{V}\frac{\partial V}{\partial c_{\alpha}} \bigg|_{p=p^{\circ}}=\frac{\partial \mu_{\alpha}}{\partial p}\bigg|_{c=c^{\circ}},
\end{align}
the baro-diffusion coefficient $D_p$ is defined as the linear combination of kinetic coefficients and partial molar volumes $\nu_{\alpha}$.  As a result of the kinetic relation (\ref{Onsager2}), 
these transport coefficients take the form
\begin{subequations}\label{DiffCoefficients}
\begin{align}
D_{+} &:={\cal L}_{++}\left(\frac{\partial \mu_+}{\partial c_+} -\beta\frac{\partial \mu_v}{\partial c_+} \right),\label{ChemicalDiffCoeff}\\
D_p& := {\cal L}_{++}\left(\nu_+- m_+\nu_0- \beta\nu_v\right).\label{BaroDiffCoeff}
\end{align}
\end{subequations}
Using the thermodynamic constraint $\sum_{\alpha \in\{+,v,0\}}\nu_{\alpha}c_{\alpha}=1$ and reformulating $\nu_0$ in terms of the cation and cation site partial molar volumes, we find (see Sec.~SI-1.2.4)
\begin{align}
    D_p& = -{\cal L}_{++}\frac{M_+}{\rho_0}\left(1-\frac{\rho}{M_+}\Delta \nu\right),\label{BaroDiffusionAllgemein}
\end{align}
where 
\begin{align}
    \Delta \nu := \nu_+-\nu_v\label{DeltaNu}
\end{align}
denotes the difference in cation and cation site partial molar volumes. Thus, due to the assumed incompressibility of the anion background lattice, their partial molar volume is negligible. In incompressible SEs without lattice deformations baro-diffusion is mainly determined by the difference between partial molar volumes of cation and cation sites, i.e., the inhomogeneity on the conduction pathway.

Besides transport in the chemical potential gradient the driving force (\ref{DrivingForce}) indicates cation transport in the electric field.
Introducing 
the electrical cation mobility by
\begin{align}
b_+ &:= {\cal L}_{++}\overline{z}F\label{Conductivity1}
\end{align}
we obtain the cation flux representation 
\begin{align}
\textbf{N}_+&= -\left(D_{+}\nabla c_+ + D_{\text{ih}}\nabla c_1
+D_p\nabla p + b_+\nabla \phi\right),\label{CationFlux4Mechanisms}
\end{align}
where
\begin{equation}
D_{\text{ih}} := {\cal L}_{++}\left(\frac{\partial \mu_+}{\partial c_1} -\beta\frac{\partial \mu_v}{\partial c_1} \right)\label{ChemicalDiffCoeffInhom}
\end{equation}
quantifies the deviation from homogeneous equilibrium distributions of cations, which is established by     
 different interacting and competing cations transport contributions. The first term in Eq. (\ref{CationFlux4Mechanisms}) describes inter-diffusion (transport in a cation concentration gradient), the second term transport with respect to the inhomogeneitiy, the third term baro-diffusion (transport in a pressure gradient), and the fourth
term drift (transport in the electric field). Thus, there are two additional terms compared to the standard model of Nernst-Planck and the phenomenological modifications \cite{Kornychev_1981, Landstorfer_2011, Danilov_2011}. In particular, the consideration of mechanical components in the free energy model leads to an additional elastic pressure-dependent term.  
Moreover, by construction all diffusion coefficients contain particle-particle interactions via the free energy and cannot be reduced to self-diffusion only.

\subsubsection{Material Model}
The next building block is the specific material model. The material model relies on a free energy functional consisting of a reference free energy density $\sum_{\alpha \in \{+,v, 0\}}c_{\alpha}\psi^{\circ}_{\alpha}$ corresponding to a configuration without external field (subsequently all quantities with superscript $\circ$ correspond to this reference configuration), the free energy density due to the entropy of mixing $\rho\psi^{\text{mix}}$, the mechanical free energy density $\rho\psi^{\text{mech}}$, and the free energy density due to polarization $\rho\psi^{\text{P}}$, i.e.,
\begin{align}
\rho\psi = \sum_{\alpha\in \{+, v,0\}}c_{\alpha}\psi^{\circ}_{\alpha}+\rho\psi^{\text{mix}}+\rho\psi^{\text{mech}}+\rho\psi^{\text{P}}.\label{FreeEnergy}
\end{align}
We model incompressible isotropic linear elastic, linear dielectric inorganic SEs (see assumption (SI-M2)). In this conceptional study, almost all contributions of free energy density are chosen as in the previous work~\cite{Braun_2015}. Only the mechanical part of the free energy density is slightly generalized, to account for size and volume effects in the lattice structure. For details see Sec.~SI-1.2.5. 
\begin{itemize}
\item 
The chemical potentials derived from this material model depend on the elastic pressure $p$. 
In the incompressible limit of diverging bulk modulus, this pressure dependency remains and the chemical potentials become
\begin{subequations}\label{IncompressibleChemicalPotentialgBYL}
\begin{alignat}{2}
\mu_{\alpha}&= \mu_{\alpha}^{\circ} +\nu_{\alpha}(p-p^{\circ})
+R\theta\ln\frac{c_{\alpha}}{c_{1}},\quad \alpha\in\{+,v\},\label{IncompressibleChemicalPotential1gBYL}\\
\mu_0&= \mu_0^{\circ} +\nu_{\alpha}(p-p^{\circ}),&&\label{IncompressibleChemicalPotential2gBYL}
\end{alignat} 
\end{subequations}
where $R$ is the universal gas constant. These chemical potentials are concrete examples of the general forms (\ref{GeneralChemicalPotentials}).
\item
The dielectric polarization is given by $\textbf{P} =\chi_{\text{SE}}\textbf{E}$, where $\chi_{\text{SE}}$ is the dielectric susceptibility.
Therefore, the  stress tensor of the linear elastic dielectric SE reads~\cite{Braun_2015}
\begin{align}
 {\cal T} = -\left(p +\frac{1}{2}\epsilon_{\text{SE}}\vert\textbf{E}\vert^2 \right) \textbf{I}+\epsilon_{\text{SE}}\textbf{E}\otimes\textbf{E},
 \end{align}
where $\epsilon_{\text{SE}}=\epsilon_0(1+\chi_{\text{SE}})$ denotes the dielectric permittivity of the SE. We notice that the magnitude of the Maxwell stress is of order ${\cal O}(\epsilon_{\text{SE}}\vert\textbf{E}\vert^2)$. Due to this quadratic dependence and the scaling with the material property $\epsilon_{\text{SE}}$, the influence of the Maxwell stress in the SCLs with pronounced electric fields is significant and cannot be neglected.
Particularly, at planar interfaces at rest the continuity of ${\cal T}$ may result in high elastic stresses.  
\end{itemize}

\subsubsection{Equations of Motion} 
The combination of the material relationships derived from the above specific material model with the transport equations completes the set of equations and provides the transport parameters, fluxes, and model equations for incompressible, inorganic SEs with uniform elastic force distribution. 

Inserting the chemical potentials (\ref{IncompressibleChemicalPotentialgBYL}) in the Eqs. (\ref{DiffCoefficients}), (\ref{ChemicalDiffCoeffInhom}) and using basic properties of the SE such as the charge symmetry, $z_+=-z_0$, and the incompressibility assumption, the transport and inhomogeneity parameter have the form
\begin{align}
D_+  &= {\cal L}_{++}R\theta\frac{\rho}{\rho_0}\frac{c_1}{(c_1-c_+)c_+},\label{ChemDiff_hom}\\
D_{\text{ih}} & = {\cal L}_{++}R\theta\frac{\rho}{\rho_0}\frac{1}{c_1-c_+},\label{Inhom_param}\\
D_p &= -{\cal L}_{++}\frac{M_+}{\rho_0}\left(1-\frac{\rho}{M_+}\Delta\nu\right),\label{BaroDiffCoeff}\\
b_+ &= {\cal L}_{++}\frac{Mc_0}{\rho_0}z_+F\label{Mobility_uniform}
\end{align}
for the three transport mechanisms in the cation flux (\ref{CationFlux4Mechanisms}) and the inhomogeneity parameter (\ref{ChemicalDiffCoeffInhom}). Each of these parameters is proportional to the kinetic factor ${\cal L}_{++}$ and a thermodynamic factor. In contrast to the baro-diffusion coefficient and the electrical cation mobility, the thermodynamic factor in the chemical diffusion coefficient and the inhomogeneity parameter depend on the different local composition. Since the chemical diffusion coefficient has to be finite, the dependence of the thermodynamic factor on the concentration implies also a concentration dependence of the kinetic factor, i.e., ${\cal L}_{++} = {\cal L}_{++}(c_+, c_1)$. For $\Delta\nu=0$, $\nabla c_1=0$, the cation diffusion flux (\ref{CationFlux4Mechanisms}) reduces to the flux derived in Braun et al.~\cite{Braun_2015}.

Merging the material relation for the dielectric polarization vector with Coulomb's law (\ref{Coulomb}) leads to Poisson's equation for the coupling between electric potential and free charge density. Similarly, we obtain the electric current density 
 \begin{align}
  \textbf{j}^{\#}&= z_{+}F\frac{\rho}{\rho_0}\textbf{N}_{+}-\partial_t \left(\chi_{\text{SE}}\nabla\phi\right).\label{Current2}
 \end{align}
Just like the cation flux (\ref{CationFlux4Mechanisms}), the free charge current density in Eq. (\ref{Current2}) differs from the usual approaches by the additional cation transport terms. The decisive difference to most other approaches, however, is that we also take the polarization current into account, which significantly contributes to the total current in the vicinity of the interfaces.

In summary, we obtain the following explicit 
equations of motion
\begin{alignat}{3}
\partial_tc_+ +\nabla \cdot \left(\frac{\rho}{\rho_0}\textbf{N}_+\right)&=0,\label{Braun1}\\
\partial_t\rho\textbf{v} +\nabla\cdot(\rho\textbf{v}\otimes\textbf{v})-\nabla p- q_F\nabla \phi&=0,\label{Braun2}\\
 \partial_t q +\nabla\cdot\textbf{j}^{\#} &=0,\label{ChargeConservation3}\\
-\nabla\cdot\left(\epsilon_{\text{SE}}\nabla\phi \right)&=q_F,\label{Braun3}
\end{alignat}
i.e., the continuity of cations, of momentum, of electric current, and the Poisson equation.

\subsection{Model Equations in Mechanical Equilibrium}
\label{Subsec_dDDP}
In this section, we evaluate the equation system (\ref{Braun1})-(\ref{Braun3}) for incompressible SEs in mechanical equilibrium, i.e., under assumption (SI-P3). In particular, we 
discuss the consequences of mechanical equilibrium on the equation system and the transport mechanisms. 

\subsubsection{Quasi-Static Momentum Equation and Driving Force}
A dimensional analysis (see Sec.~SI-1.3.1) shows that, we can neglect dynamic and convective contributions in (\ref{Braun2}) and obtain a quasi-static relation for the pressure gradients
\begin{align}
 \nabla p =-q_F\nabla\phi.\label{ForceBalance}
\end{align}
This approximation describes thermodynamic processes in which mechanical equilibrium is reached much faster than the diffusion processes in the system. With other words, on the time scale relevant for most battery applications the incompressible, elastic monocrystalline SEs is in mechanical equilibrium.  

In mechanical equilibrium the elastic
pressure is a dependent quantity determined up to a constant by Eq. (\ref{ForceBalance}). This coupling of the elastic and the electrostatic forces allows us to eliminate one degree of freedom and to rewrite the chemical potential gradients of mobile species (\ref{chemicalPotentialGradientGeneral}) to 
\begin{align}
\nabla \mu_{\alpha} &=\frac{\partial \mu_{\alpha}}{\partial c_{+}}\nabla c_{\alpha} + \frac{\partial \mu_{\alpha}}{\partial c_{1}}\nabla c_{1}  - \nu_{\alpha}q_F\nabla \phi. \label{ChemicalPotentialGradientGeneral2}
\end{align} 
Hence, it leads to a dependence of the chemical potential gradients on the Coulomb force density. 
With Poisson's equation (\ref{Braun3}) this electrostatic force in Eq. (\ref{ChemicalPotentialGradientGeneral2}) can be replaced by a second order polynomial of the potential gradient, which then also enters the driving force (\ref{DrivingForce}). For details see Sec.~SI-1.3.2.\footnote{Subsequently, quantities in mechanical equilibrium are indicated by a hat.}

\subsubsection{Drift-Diffusion-Flux and Transport Coefficients}
Classical Nernst-Planck theory~\cite{Nernst_1889, Planck_1890a, Planck_1890b} as well as for instance the phenomenologically modified approaches\cite{Kornychev_1981, Danilov_2011} considers cation transport as diffusion and conduction. In SCLs coupling to electrostatic forces provides an additional 
conductivity mechanisms. By defining the free charge dependent cation conductivity
\begin{align}
\hat{\sigma}^{\#}: =z_+F\frac{\rho}{\rho_0}(b_+ - D_p q_F)\label{DefConductivity}
\end{align}
which relates the potential gradient $\nabla \phi$ to the electrical current  (\ref{PolarizationCurrent}),
we transform the cation diffusion flux (\ref{CationFlux4Mechanisms}) to the form
\begin{align} 
\hat{\textbf{N}}_+&=- \left( D_{+}\nabla c_+ +  D_{\text{ih}}\nabla c_1
+\frac{\hat{\sigma}^{\#}}{z_+F}\frac{\rho_0}{\rho}\nabla \phi\right).\label{driftDiffusionFlux}
\end{align}
For homogeneous SEs, the cation flux (\ref{driftDiffusionFlux}) structurally reduces to a generalized drift-diffusion flux, which contains only the two transport mechanisms chemical diffusion and electro-migration with charge and concentration dependent transport coefficients. Their specific form depends on the material laws i.e., the underlying free energy. However, we can also derive principle properties of the transport parameters and analytical relations between them without specifying the free energy.

First, we notice that in the charge-neutral bulk volume and in SEs with negligible baro-diffusion process Eq. (\ref{DefConductivity}) reduces to the standard relation between conductivity and electrical cation mobility. But in SCLs the concentration dependent conductivity  $\hat{\sigma}^{\#} =\hat{\sigma}^{\#}(c_+,c_1)$  deviates from this relation.
By inserting the baro-diffusion coefficient (\ref{BaroDiffCoeff}), the cation mobility (\ref{Mobility_uniform}), and reformulating the free charge density the conductivity becomes (for details see Sec.~SI-1.3.3)
\begin{align}
\begin{split}
\hat{\sigma}^{\#} &= (z_+F)^2 {\cal L}_{++}\left(\frac{\rho}{\rho_0}\right)^2\left(1-\left(c_+-c_0\right)\Delta\nu\right)\\
&= (z_+F)^2 {\cal L}_{++}^{\#}\left(1-\left(c_+-c_0\right)\Delta\nu\right),
\end{split}
\end{align}
where ${\cal L}_{++}^{\#}$ denotes the Onsager coefficient in a lattice-fixed frame. 
For $\Delta \nu=0$, i.e., for homogeneous conduction pathways, this conductivity decouples from the space charge ---except for the concentration dependence of the kinetic factor ${\cal L}_{++}^{\#} = {\cal L}_{++}^{\#}(c_+, c_1)$---and the model reduces to the earlier model of Kornychev and Vorotyntsev~\cite{Kornychev_1981}.  We interpret this to imply that for homogeneous lattice distances without molar volume differences and lattice deformation, the lattice completely absorbs the Maxwell stress generated by the SCLs via the pressure.

Second, we study non-ideality, by rewriting the thermodynamic factors in the chemical diffusion $D_{+}=D_{+}(c_+, c_1)$ in terms of thermodynamic  factors, respectively, non-ideality factors. Using the standard representation $\mu_{\alpha}=\mu_{\alpha}^{\circ}+R\theta\ln (\gamma_{\alpha} c_{\alpha})$ of the chemical potentials in terms of chemical activity coefficients $\gamma_{\alpha}$, which describe non-ideal behaviour, and application of chain rule leads to
 \begin{align}
D_{+} &=\frac{{\cal L}_{++}R\theta}{c_+} \left(f_{++} - \beta f_{+v}\right),\label{Diffusivity}
\end{align} 
with non-ideality factors $f_{\alpha\beta}=\delta_{\alpha\beta}+\partial \ln (\gamma_{\alpha})/\partial \ln (c_{\beta})$ and $\delta_{\alpha\beta}$ the Kronecker delta. 
The cross-species non-ideality factors take account of the interaction between cations and vacant cation sites. Thus, the cation inter-diffusion
depends on the intra-species forces as well as inter-species
forces. This diffusion coefficient is related to the diffusion coefficient in the lattice-fixed frame by $D_{+}= \rho_0/\rho D_{+}^{\#}$.
 
Finally, a generalized Nernst-Einstein relation can be formulated combining the expression for interdiffusion coefficient and conductivity as 
\begin{align} \label{Nernst-Einstein_b}
\hat{\sigma}^{\#}  = D_{+} \frac{c_+ (z_+ F)^2}{R \theta} \frac{1-\left(c_+-c_0\right)\Delta\nu}{f_{++} - \beta f_{+v}}\frac{\rho}{\rho_0}.
\end{align}
The standard Nernst-Einstein relation is obtained for homogeneous conduction pathways with constant pressure 
and non-ideality factors identical to one. 

\subsection{Initial and Boundary Conditions}
\label{Subsec_IC}
In order to close the systems of equations given in the previous sections, it is necessary to state appropriate initial and boundary conditions of the systems. For comparison with the experimental findings in Refs.\cite{Yamamoto2010, Hirayama_2017, Aizawa_2017}, we polarize SEs between ion-blocking metal electrodes (see Sec.~SI-1).

Before the dielectric SE is polarized, the cations are distributed  throughout the lattice such that no electro-quasi-static potential is present, and the elastic pressure in the SE equals the ambient pressure. Since we assume dielectric materials, initially the charge density is zero. This corresponds to the conditions
\begin{equation}
c_+(0,\textbf{x})= c_+^I, p(0, \textbf{x})=p^{\circ}, q(0, \textbf{x})=0, \phi(0,\textbf{x})=0,
\end{equation}
where the constant initial distribution $c_+^I$ may be inhomogeneous and depends on the crystal structure and we chose the ambient pressure as reference pressure.
The potential drop $U$ in the electrolyte induced by the polarization experiment (see assumptions (SI-S1) and (SI-S2)), is represented by Dirichlet boundary conditions at the electrodes 
\begin{align}
\phi = U, \quad  \text{on }\Gamma_{\text{C}},\qquad\text{and}\qquad\phi=0,  \quad \text{on } \Gamma_{\text{A}}.\label{BC2}
\end{align} 
As a further consequence of assumptions (SI-S1), (SI-S2), and (SI-P4), we impose no-flux boundary conditions for the cation on the blocking electrode-SE interfaces
\begin{alignat}{2}
\textbf{N}_+^{\#}\cdot\mbox{\boldmath$\nu$}&=0,\quad &\text{on }& \Gamma_{i}, \quad i\in\{\text{A,C}\}.\label{BC3}
\end{alignat}

\begin{figure}
\begin{center}
\includegraphics[width=0.5\textwidth]{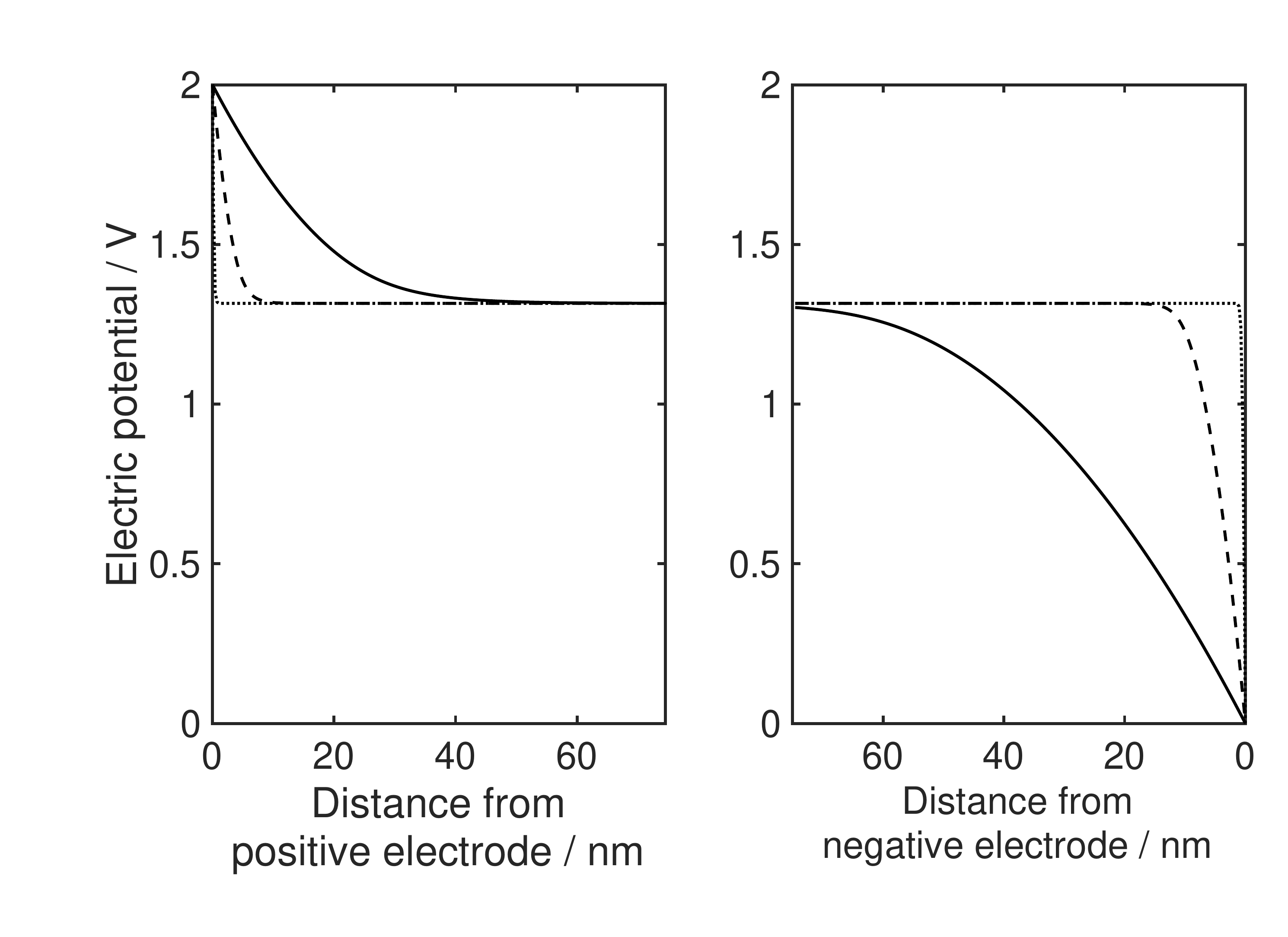}
\includegraphics[width=0.5\textwidth]{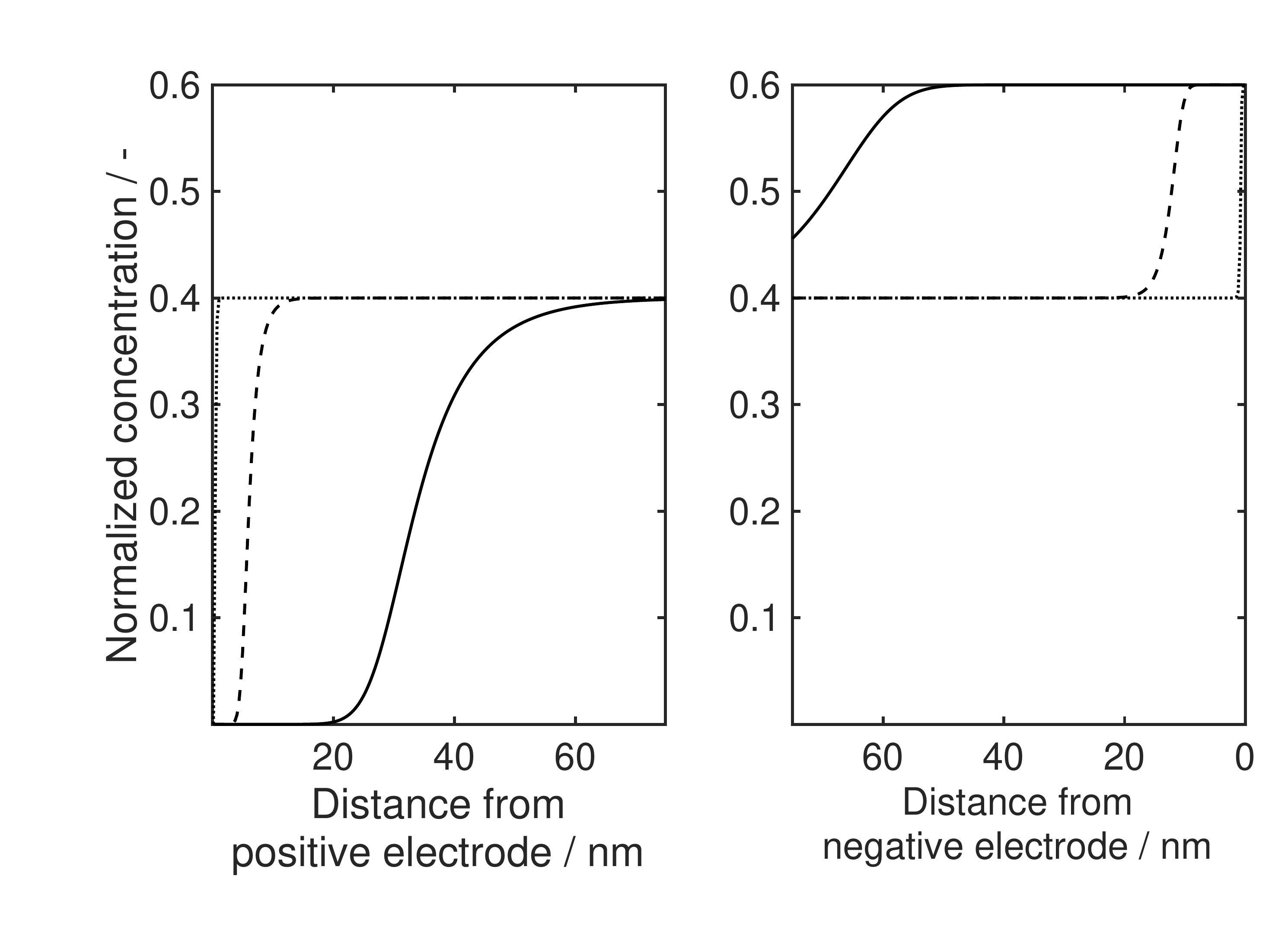}
\includegraphics[width=0.5\textwidth]{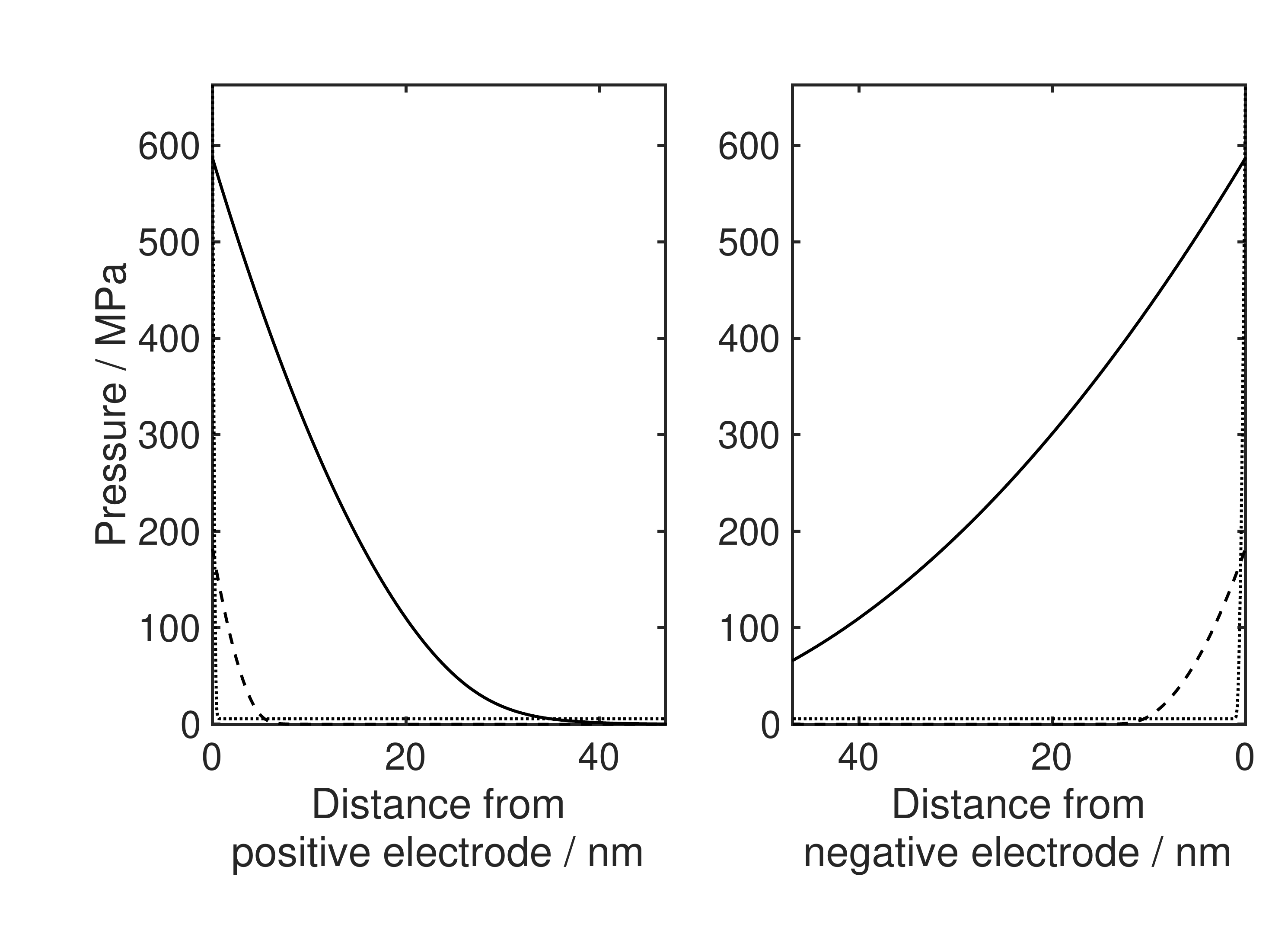}
\caption{State variable distributions next to the SE-electrode-interfaces in 2400 nm thick LLTO (solid), LATP (dashed), and LiPON (dotted) in equilibrium established after polarized to 2~V. Material parameters have been chosen according to Tab.~SI-1.}
\label{Fig_matVergleichPotential}
\end{center}
\end{figure}

\subsection{Thermodynamic Equilibrium}
\label{Subsec_EQ}
We conclude this section by investigating a SE in mutual thermodynamic equilibrium. That is, the SE is simultaneously in thermal, mechanical, and chemical equilibrium and there are no net flows of matter or energy, i.e., $\textbf{N}_+=\hat{\textbf{N}}_+=\hat{\textbf{N}}_+^{\#}=0$. This vanishing flux condition is equivalent to vanishing driving force. 

Following a concept introduced by Kornychev and Vorotyntsev~\cite{Kornychev_1981}, we first show that
steady state properties are strongly affected by the choice of the concrete material model and that equilibrium is not determined by the individual transport parameters but by the ratio of conductivity and diffusivity. This ratio enables us to formulate the vanishing driving force condition compactly, which is the key to our analytical results for SCLs in incompressible homogeneous SEs developed in Sec.~\ref{Subsec_EqResults}. 
Then, based on the effective driving force introduced this way the generalized Poisson-Boltzmann equations characterizing the equilibrium distributions of the state variables are established and discussed. 

\subsubsection{Equilibrium Characterisation}
We first consider the ratio between conductivity and diffusivity\cite{Kornychev_1981}
 \begin{align}
\hat{f}^{\#}(c_+, c_1) &:=\frac{\hat{\sigma}^{\#}(c_+, c_1)}{z_+F D^{\#}_+(c_+, c_1)},\label{Ratio1}
\end{align}
which determines in homogeneous systems their equilibrium properties by the relation
\begin{align}
-\nabla \phi =\frac{1}{\hat{f}^{\#}(c_+, c_1)}\nabla c_+.\label{EqProperty}
\end{align}
Using the canonical expression of the chemical diffusion coefficient (\ref{ChemicalDiffCoeff}) as linear combinations of kinetic factors and thermodynamic factors and the ion conductivity (\ref{Conductivity1}) the ratio (\ref{Ratio1}) becomes  
 \begin{align}
\hat{f}^{\#}(c_+, c_1) &=\frac{z_+F\left(1-(c_+-c_0)\Delta\nu\right)}{\frac{\partial \mu_+}{\partial c_+} -\beta\frac{\partial \mu_v}{\partial c_+}}.\label{Ratio}
\end{align}
In principle, due to the coupling between charge and elastic force in SCLs and the additional conduction mechanism induced thereby, this ratio depends not only on the thermodynamic factors but also on the difference in the the molar partial volumes on the conduction pathway.
But in incompressible SE without size effects the ratio is identical to that stated by Kornychev and Vorotyntsev~\cite{Kornychev_1981}, who studied cation and potential distributions in SEs without mechanical effects. Therefore, we obtain in this case equilibrium cation and potential distributions qualitatively similar them.

Next, we introduce the effective chemical potential for the population on the conduction pathway based on the equilibrium characterisation (\ref{EqProperty}) 
\begin{align}
\hat{\mu}:=\int\frac{1}{\hat{f}^{\#}(c_+, c_1)}\, dc_+ +\hat{\mu}^{\circ}.\label{chemPot}
\end{align}
With this setting thermodynamic equilibrium is characterized by the condition
\begin{align}
\nabla\hat{\varphi}:=\nabla\hat{\mu} +z_+F\nabla\phi = 0.\label{drivingForce}
\end{align}
Integration of the vanishing driving force (\ref{drivingForce}) gives that in equilibrium the effective electrochemical potential $\hat{\varphi}=\hat{\mu}+z_+F\phi$ is constant. 
This electrochemical potential allows us to define a scalar $\Phi := (\hat{\varphi}-\hat{\mu}^{\circ})/(z_+F)$, which acts like a quasi Fermi (electrochemical) potential.

\subsubsection{Generalized Poisson-Boltzmann Equation}
To study thermodynamic equilibrium, we regard the model equations
 in the 1D domain $(0,L_{\text{SE}})$. In thermodynamic equilibrium the one dimensional boundary value problem consists of the vanishing driving force condition (\ref{drivingForce}), the Poisson equation (\ref{Braun3}), the incompressibility constraints, and the boundary conditions (\ref{BC2}). To guarantee global charge neutrality and mass conservation, which is no longer ensured by the cation continuity equation (\ref{Braun1}), the system of equations is supplemented with the two integral constraints
\begin{alignat}{3}
\int_{0}^{L_{\text{SE}}}q_F(x)\, dx &=0\qquad\text{and}\qquad &\int_{0}^{L_{\text{SE}}} c_+(x)\, dx& =\overline{c}_+\label{MassConservation},
\end{alignat}
with $\overline{c}_+ ={\cal M}_+/ M_+$.
In concrete terms, the following effective electrochemical potential results for the material model
\begin{align}
\begin{split}
\hat{\varphi}&=\hat{\mu}^{\circ}+ R\theta\ln\left( \frac{c_+}{c_1-c_+}\right)\\&+ z_+F\left(\phi -\int (c_+-c_0)\Delta\nu\partial_x\phi\, dx\right).
\end{split}
\end{align}
Rewriting this effective electrochemical potentials in terms of the cation concentration and using the definition of the quasi Fermi potential gives
the cation distribution 
\begin{align}
c_+&=\frac{c_+^{\text{max}}}{1+\exp\left(\frac{z_+F}{R\theta}\left(\phi-\Phi-\int (c_+-c_0)\Delta\nu\partial_x\phi\, dx\right)\right)}.\label{Eq_CationDistribution2}
\end{align}
The cation distribution  (\ref{Eq_CationDistribution2}) is given by a logistic sigmoid function of the electric potential and its derivative.  The coupling of mechanics and diffusion is reflected by the second order polynomial in $\partial_x\phi$. In homogeneous SE with equal partial molar volumes, i.e., for $\Delta\nu=0$, the distribution has the form of a Fermi-Dirac distribution, which simply expresses the fact that there are a limited number of sites in a finite crystal and that these sites can either be filled or empty.  
The maximum value $c_+^{\text{max}}$ of the function (\ref{Eq_CationDistribution2}) is given by the concentration of mobile cations sites $c_1$ describing fully saturated cation sites times the positive integration constant $C_+$, which may be identified by the mass conservation side condition (\ref{MassConservation}). Irrespective of how the SE is polarized, the local cation concentration never exceeds the maximum value of  cation lattice sites on the conduction pathways.

Inserting the cation distribution (\ref{Eq_CationDistribution2}) in Poisson's equation (\ref{Braun3}) yields a generalized Poisson-Boltzmann (gPB) equation governing the electro-quasi-static potential $\phi$ in equilibrium for given boundary values and constraints 
\begin{align}
-\epsilon_{\text{SE}}\partial_{xx}  \phi=  \frac{ z_+Fc_+^{\text{max}}}{1+\exp\left(\frac{z_+F}{R\theta}\left(\phi-\Phi-\int (c_+-c_0)\Delta\nu\partial_x\phi\, dx\right)\right)} +z_0Fc_0.\label{PB2}
\end{align}
In incompressible SE without different size effects and lattice distances, i.e., for $\Delta\nu=0$, we obtain a gPB equation similar to 
Kornychev and Vorotyntsev~\cite{Kornychev_1981}. But even in this case, the gPB equation differs from the classical Gouy-Chapman Poisson-Boltzmann equation~\cite{Chapman_1913}, which leads to extreme, non-physical values of the cation concentrations in the SCLs. 
In contrast to the exponential cation distribution in space predicted by these approaches, this gives a hyperbolic distribution with respect to the electrical energy of the SE induced by its polarization. This results in a non-exponential distributions of the cations in space to shield the electric field. Therefore, our subsequent numerical results also differ from the work on SEs, which phenomenologically modify the classical Gouy-Chapmann Poisson equilibrium approach by additional phenomena such as grain boundaries\cite{GoebelI_2014, GoebelII_2014, Maier_2017} and Coulomb's interactions~\cite{deKlerk_2018}.

On the other hand, the equation also differs from the modified Poisson-Boltzmann equations for ionic and liquid electrolytes with volume constraints~\cite{Kornychev_2007,Bazant_2009} by the free charge density as an asymmetric logistic sigmoid functional of the electric potential as function of space. The asymmetry of the free charge density is a result of the specific characteristic of SEs---fixed anions and  cation site hopping transport mechanism of the cations~\cite{Braun_2015}. 
Since  
\begin{align}
q_{F}\in [-z_0 Fc_0, z_+Fc_+^{\text{max}}-z_0Fc_0],
\end{align}
the free charge density $q_F$ is symmetric if and only if  $c_1  = 2c_0$. As we will see below, these two differences to the previous work mentioned above provide the basis for our qualitative results.

\section{Simulation Details}
\label{Sec_SimulationDetails}
For the simulations, we implement our model in Matlab. The computational details for the simulation results shown in the next section are given in the supporting information in Sec.~SI-2. On the one hand a short overview of the numerical methods used is given there and on the other hand the parameters used are specified and discussed. In particular, this includes a brief discussion of the controversial magnitude of dielectric susceptibility in polycrystalline materials.

In our simulations, we consider three different types of SEs as examples: the perovskite Li$_{0.5}$La$_{0.5}$TiO$_3$ (LLTO), Li$_{1.3}$Al$_{0.3}$Ti$_{1.7}$(PO$_4$)$_3$ (LATP) with a (Na$^+$) super ion conductor (NaSICON) type structure and amorphous LiPO$_{4}$ (LiPON). This selection allows us to compare the results with our previous work~\cite{Braun_2015}, as well as other modelling work~\cite{deKlerk_2018} and experimental results~\cite{Yamamoto2010, Hirayama_2017, Aizawa_2017, Nomura_2019}. For simplicity, we assume that cations and cation sites have the same
molar volume and that the equilibrium distribution of cations is homogeneous.
Like the material selection, the geometry and experiment dependent parameters were chosen with regard to the comparability to these works.

\section{Results and Discussion}
\label{Sec_EquilibriumResults}
In this section we present both analytical and simulation results for polarized homogeneous SEs without size effects on the conduction pathway ($\Delta\nu=0, \nabla c_1=0$) sandwiched between two ion-blocking electrodes. To begin, in Sec.~\ref{Subsec_EqResults} the equilibrium properties of SCLs
are studied. A low temperature approximation allows us to explicitly calculate the equilibrium width of the SCLs, which exceeds the classical Debye screening length by at least one order of magnitude. Then, in Sec.~\ref{Subsec_Dynamik} the process of SCL formation is revealed by numerical simulations. Special focus lies on the
the different components of the current response to polarization. Detailed derivations of the analytical results are given in the supporting information in Sec.~SI-3. 

\subsection{Space-Charge-Layers In Equilibrium}
\label{Subsec_EqResults}
First, we examine the material parameter dependence of the SCLs and present an analytical expression for their widths in equilibrium as well as 
for the bulk electrolyte potential level, to which they converge.
Our results match to the experimental characterizations by electron holography\cite{Yamamoto2010, Hirayama_2017, Aizawa_2017, Nomura_2019} and explain the observed asymmetric potentials. Furthermore, we calculate the differential SCL capacity.

\begin{figure}
\begin{center}
\includegraphics[width=0.5\textwidth]{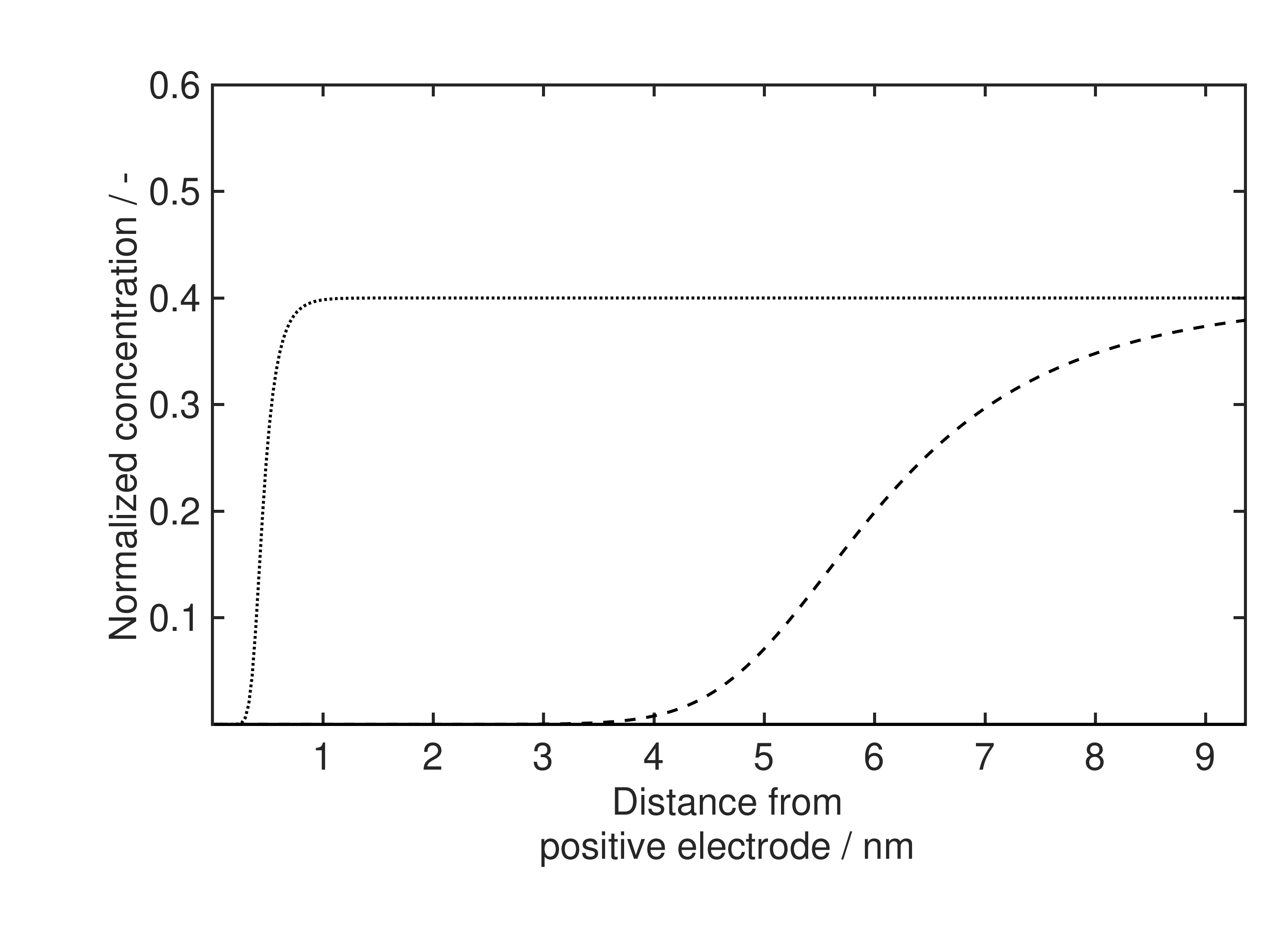}
\caption{Cation distributions next to the positive electrode-SE interfaces in 2400 nm thick LATP (dashed), LiPON (dotted) in equilibrium established after polarized to 2~V for material parameters chosen according to Tab.~SI-1.}
\label{Fig_matVergleichPotentialZoom}
\end{center}
\end{figure}

\subsubsection{Equilibrium Distribution of State Variables}
\label{Sec_NumericEquilibrium}
We illustrate the basic characteristics of state variable distributions established in equilibrium by numerical results for polarized 2400 nm-thick LLTO, LATP, and LiPON layers. 
As discussed in the supporting information in Sec.~SI-2.1, the accumulation of charges at grain boundaries leads to macroscopic polarization. Maxwell-Garnett effective medium theory like considerations suggest that under finite current conditions the relevant dielectric suceptibility in polycrystalline SEs is rather close to the high low frequency magnitude than to the lower high frequency magnitude~\cite{Hanai_1960, Neumann_2019}. To account for this theoretical possibilities, 
we choose low frequency and high frequency values of the dielectric function, respectively, to explore the possible impact on SCLs. 
The equilibrium distributions near the electrode interfaces in the different materials are compared in Fig.~\ref{Fig_matVergleichPotential}. 
Fig.~\ref{Fig_matVergleichPotentialZoom} shows the cation distribution at the positive electrode-SE interface spatially in LATP (dashed) and LiPON (dotted) in more detail. 
We note from both Fig.~\ref{Fig_matVergleichPotential} and Fig.~\ref{Fig_matVergleichPotentialZoom} that the potential and concentration profiles obtained by the simulations have the same qualitative characteristics: 
\begin{itemize}
\item
Fig.~\ref{Fig_matVergleichPotential} (top) shows that for the three chosen simulation parameters the potential distributions in the materials 
exhibit a potential drop of $\Delta \phi_{\text{C}}$=-0.7~V at the SE-positive-electrode interface, a bulk plateau level of $\phi_{\text{bulk}}$= 1.3~V, and apparently a potential drop of $\Delta \phi_{\text{A}}$=-1.3~V at the SE-negative-electrode interface.
\item
The cation concentration exhibits an accumulation zone at the electrode with low potential and a depletion zone at the higher potential electrode which are connected by the charge neutral zone in the bulk. Different from liquid electrolytes the depletion and accumulation zones are asymmetric (Fig.~\ref{Fig_matVergleichPotential}) except for singular values of the anion concentration in the charge neutral bulk~\cite{Braun_2015}. The calculated spatial variation of the concentration is clearly 
non-exponential. The plateau for the concentration close to the interfaces seen in Fig.~  
\ref{Fig_matVergleichPotential} (middle) up to the turning point of the sigmoid function corresponds to a quadratic variation of the potential in 
Fig. \ref{Fig_matVergleichPotential} (top). 
\item The spatial extent of the cation degradation and accumulation zones as well as the electrical potential scales with the Debye length~\cite{Debye_1928}
$\lambda_{\text{D}}=\sqrt{\epsilon_{\text{SE}}R\theta/(c^{R}F^2)}$, i.e., with $\sqrt{\varepsilon_{\text{SE}}}$. This is demonstrated in Fig ~\ref{Fig_matVergleichPotentialZoom}. But the absolute value of the depletion zone at the positive electrode-SE interface is much larger than the Debye length (about 10 times as large in the case of LLTO with 30~nm, as in Fig.~\ref{Fig_matVergleichPotential} (middle) vs. $\lambda_D \approx 3$~nm).   
\end{itemize}

As shown in Fig.~\ref{Fig_matVergleichPotential} (bottom), in the SCLs we observe local elastic forces in the range of MPa. The potential distribution in Fig.~\ref{Fig_matVergleichPotential} (top) indicates that the smaller the SCLs, the greater the local electric fields. Closed to the positive electrode-SE interface, we identify in LLTO, LATP, and LiPON electric fields in the range of 10$^{7}$, 10$^{8}$, and 10$^{9}$ Vm$^{-1}$. Therefore, for materials with similar mass and mass density, such as LLTO (solid) and LiPON (dashed), the coupling of the elastic and electrostatic forces leads to an increase in the elastic forces as the SCL width decreases (see Fig.~\ref{Fig_matVergleichPotential} bottom).  In contrast, the local electric field in the heavier super-ionic conductor LATP causes significantly lower electric forces and the lattice requires lower elastic forces (dashed) to absorb them.

\begin{figure}[hbt]
\begin{center}
\includegraphics[width=0.35\textwidth]{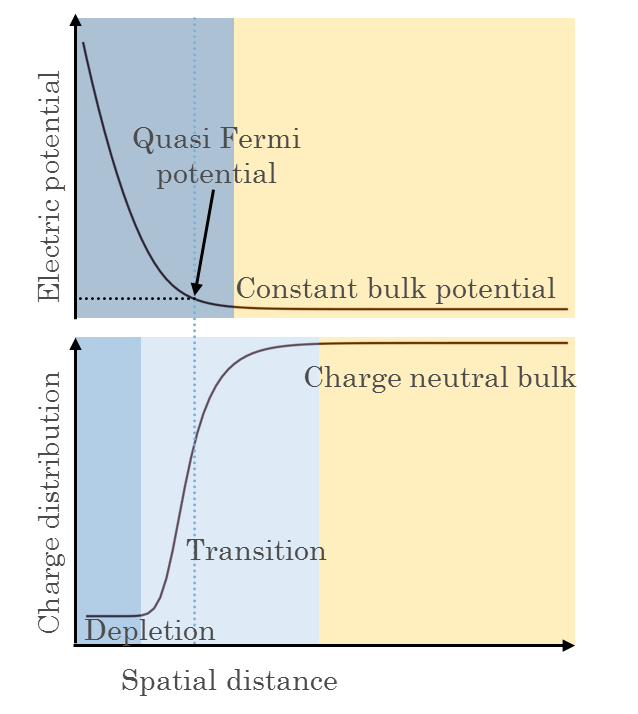}
\caption{Schematic illustration of the potential (top) and the cation distributions (bottom) formed at the positive electrode-SE interface after polarization. Blue areas indicate space charge layer while yellow areas represent bulk regions. In the vicinity of the interface the cations are depopulated forming a depletion zone (bottom dark blue area). This is followed by a transition zone (bottom light blue area). Both the depletion and the transition zone together form the ionic SCLs.}
\label{Fig_SchematicSCLs}
\end{center}
\end{figure}

\subsubsection{The Quasi Fermi Potential}
Although the ionic charge distribution and the potential distribution have a one to one relation via the Poisson equation, the characteristic shape of both is quite different (see Fig.~\ref{Fig_SchematicSCLs}). The width of both layers appear to be different, since the ionic charge distribution is coded in the curvature of the potential distribution and the transition from a linear to a curved shape of the potential, e.g., in electron holography measurements~\cite{Aizawa_2017, Nomura_2018}, is not easily detected. 
 The quasi Fermi potential (doted lines) scales with position of zero curvature in the ion distribution or the position of $c_+=c^{\text{bulk}}_+/2$ in the depletion zone or $c_+= (c^{\text{max}}_++c^{\text{bulk}}_+)/2$ in the accumulation zone, respectively.  

To investigate this in more detail, we explicitly calculate the quasi Fermi potential via the chemical potential $\hat{\mu}_L=\hat{\mu}(0)$ at the positive-electrode-SE interface. The derivation uses the global charge neutrality condition (\ref{MassConservation}) and the fact that the electrochemical potential $\hat{\varphi}$ is constant in equilibrium. Details are given in Sec.~SI-3.1. We find that the quasi Fermi potential is a function of applied voltage and material composition
\begin{align}
\Phi = U + \frac{R\theta}{z_+F} \ln \left(\frac{\exp\left(\frac{c_0}{c_1}z_+FU \right)-1}{\exp(z_+FU)-\exp(\frac{c_0}{c_1}z_+FU)}\right).
\label{FermiPotential}
\end{align}
The bulk potential can also be calculated from the constant electrochemical potential $\hat{\varphi} = \hat{\mu}_{\text{bulk}} + z_+F\phi_{\text{bulk}}$ using charge neutrality 
\begin{align}
\phi_{\text{bulk}} &=\Phi -\frac{\hat{\mu}_{\text{bulk}}-\hat{\mu}^{\circ}}{z_+F} . \label{BulkPotential1}
\end{align}
From Eq.~(\ref{FermiPotential}) and Eq.~(\ref{BulkPotential1}) it is evident, that neither the quasi Fermi potential nor the bulk potential depend on the dielectric properties of the material. This results in the fixed bulk plateau for given composition already observed in Fig.~\ref{Fig_matVergleichPotential} (top). At the same time, it also becomes clear that, by changing the number of  cation sites on the conduction pathway the position of the quasi Fermi potential and thus the bulk potential level, and, especially, the potential drops in the SCLs can be changed.

\begin{figure*}[htbp]
\includegraphics[width=0.5\textwidth]{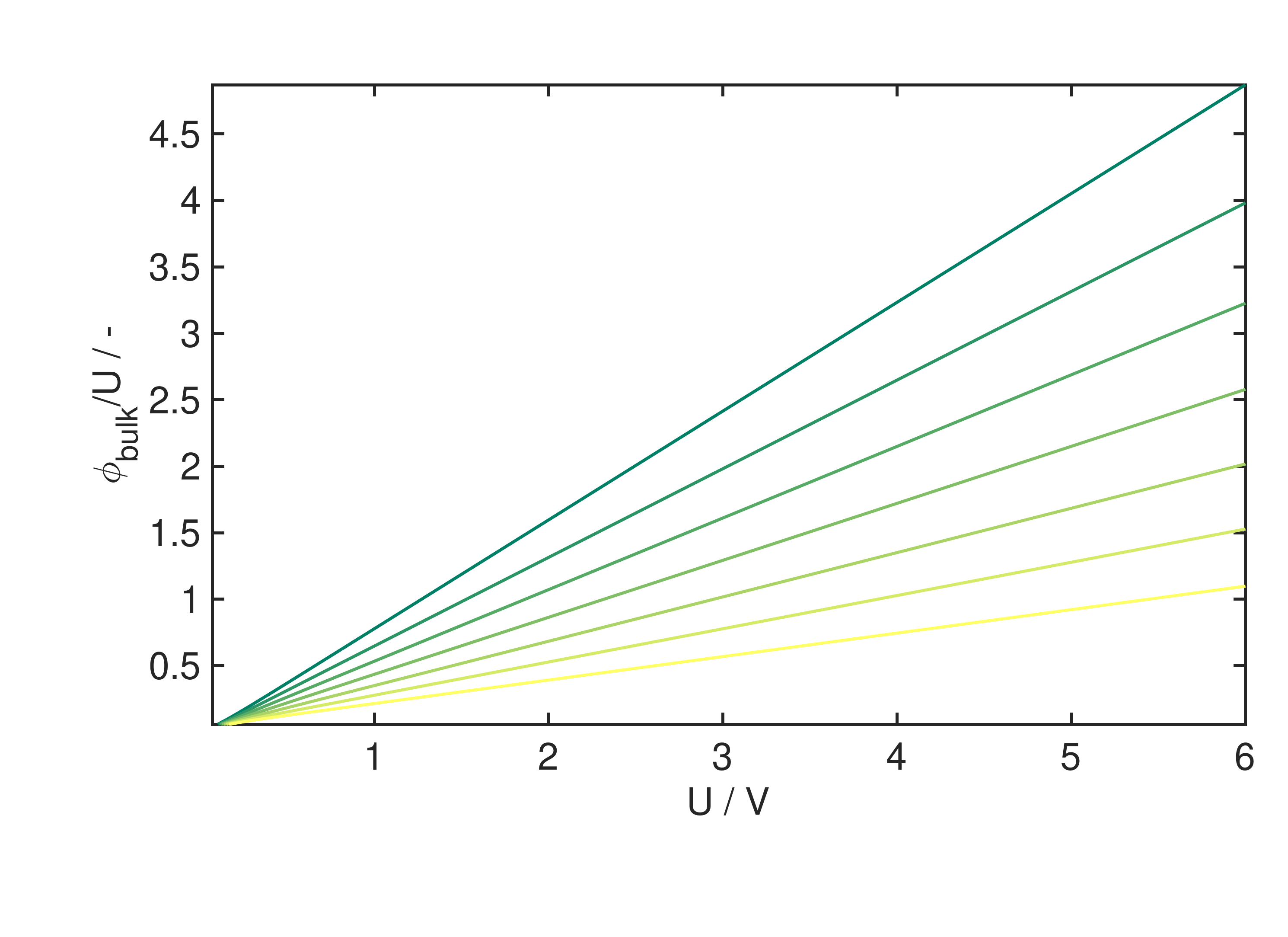}\includegraphics[width=0.5\textwidth]{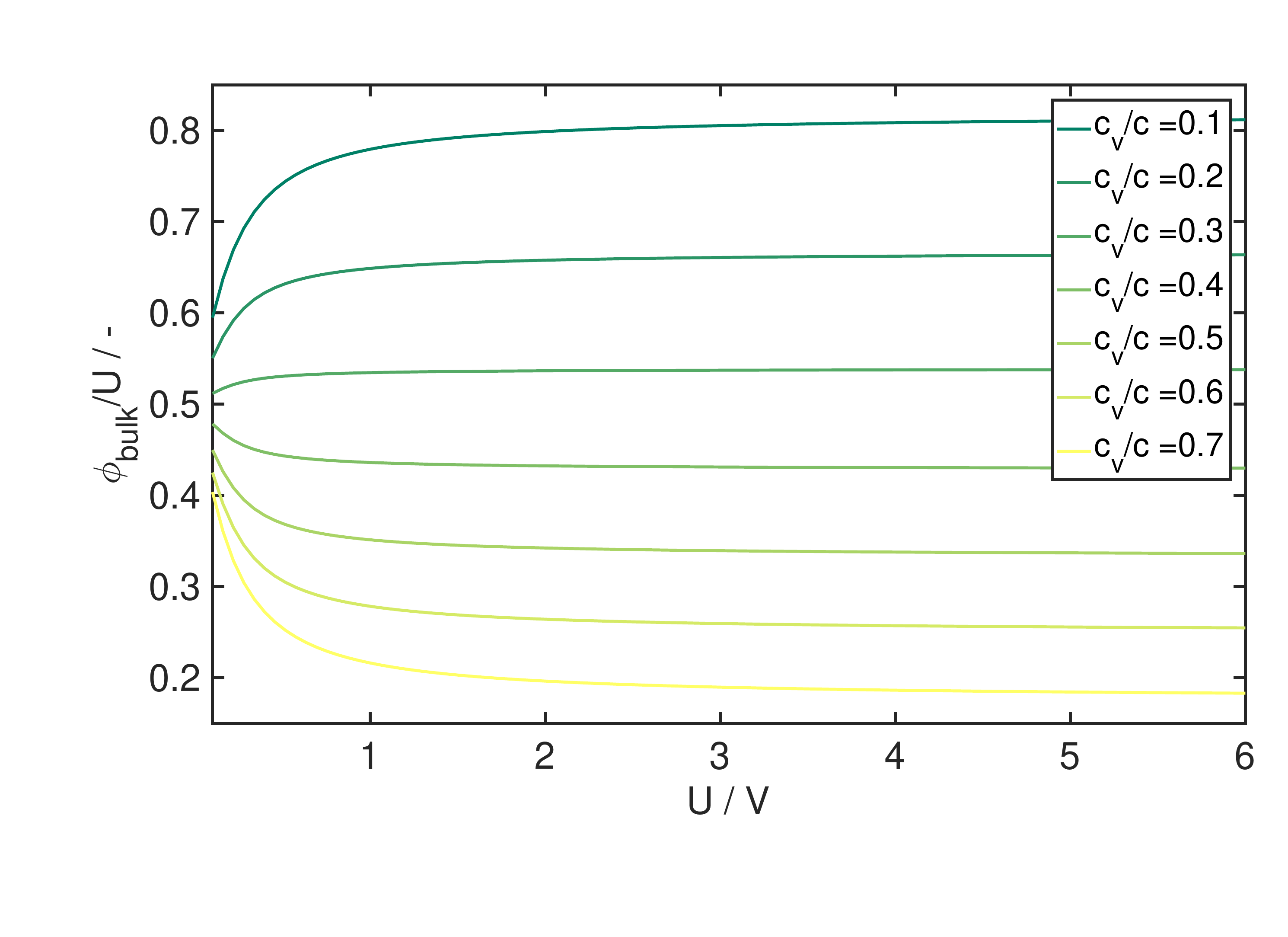}
\caption{Absolute height of the bulk potential plateau (left) and normalized value (right) as functions of applied potential difference $U$. Both are shown in color gradients for variations of dimensionless  defects.}
\label{Fig_BulkPotential}
\begin{center}
\includegraphics[width=0.5\textwidth]{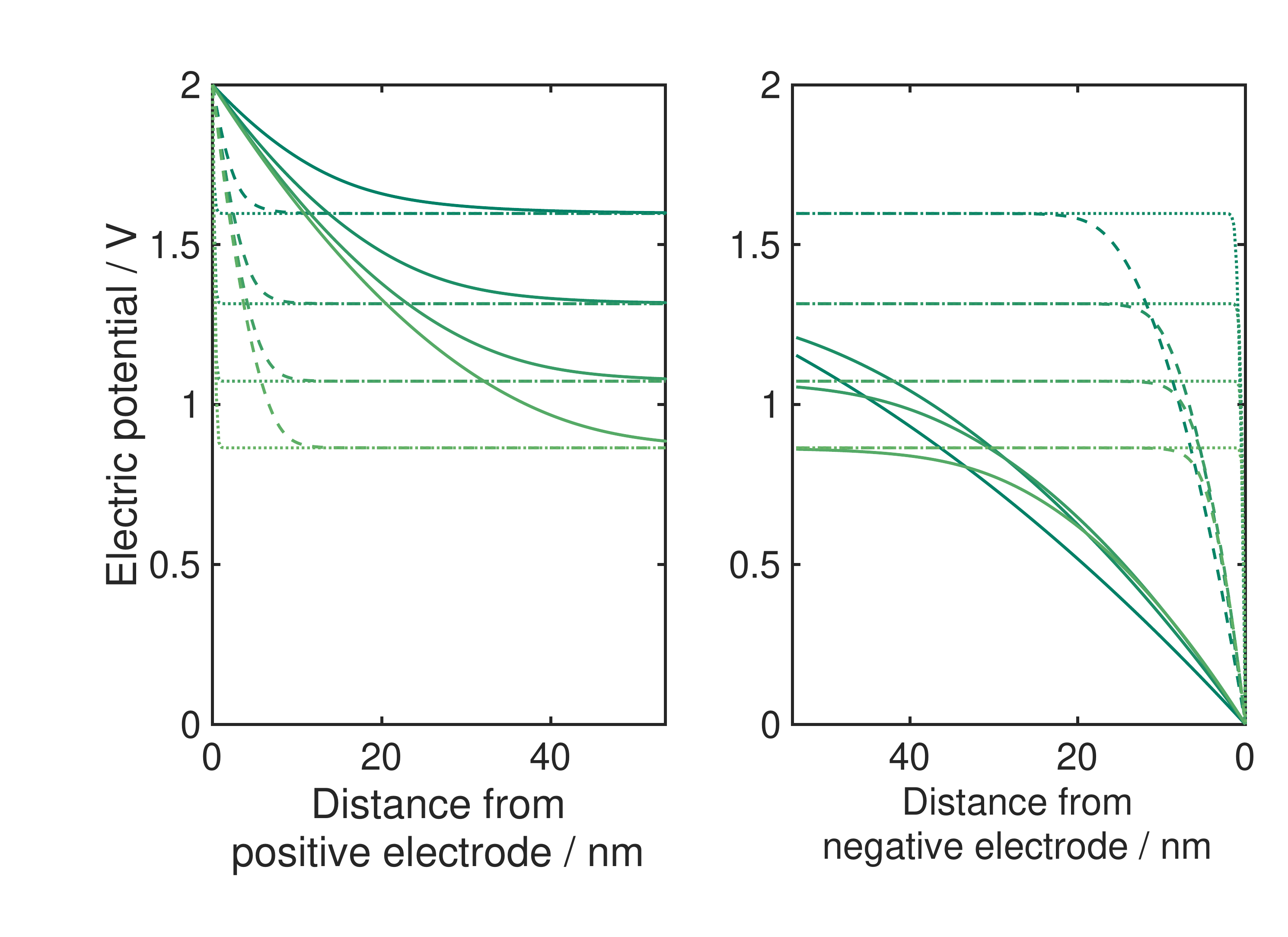}\includegraphics[width=0.5\textwidth]{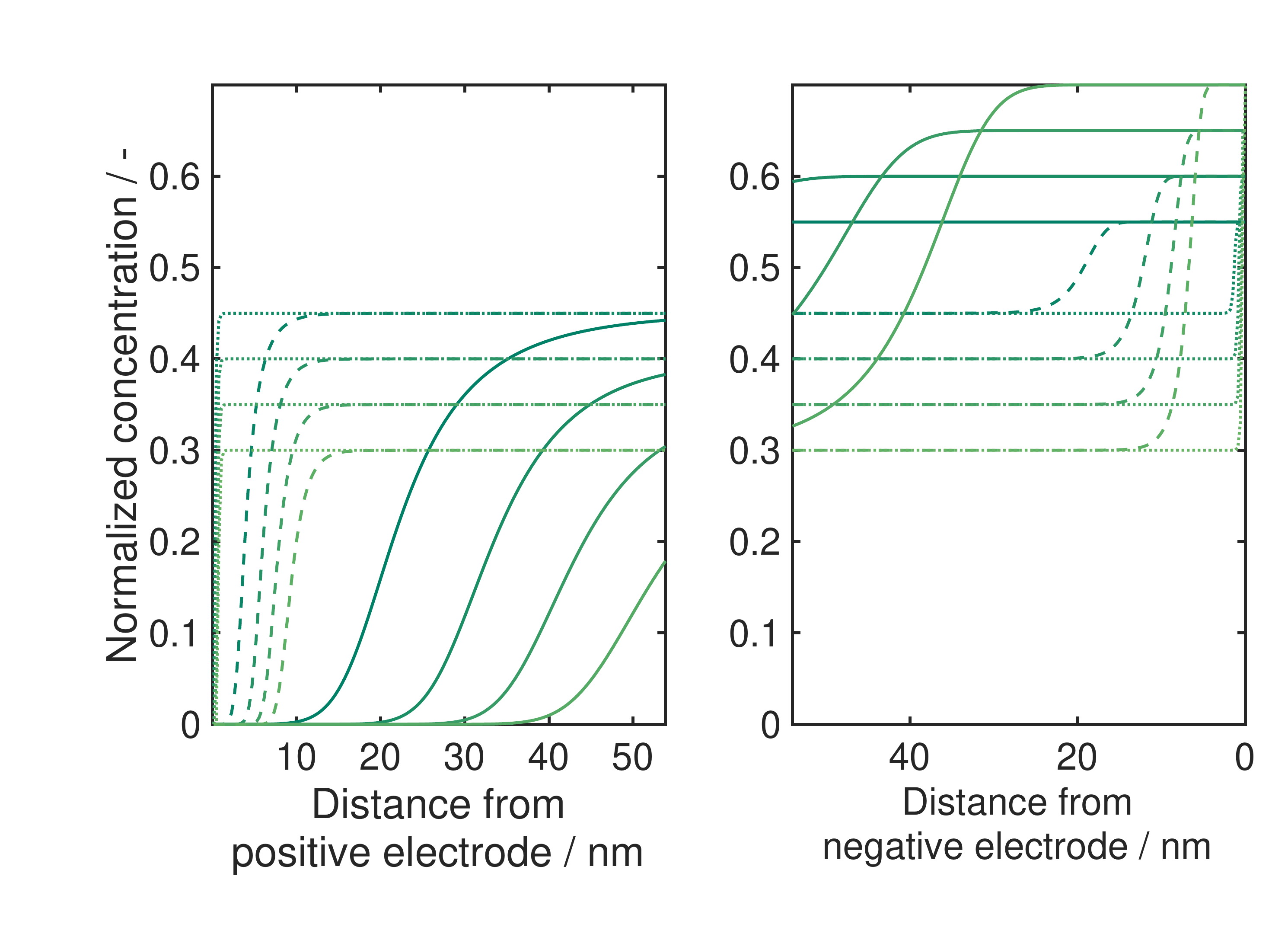}
\caption{Equilibrium potential (left) and cation concentration (right) profiles near the electrode interfaces 2400 nm-thick LLTO (solid), LATP (dashed), and LiPON (dotted) are shown in color gradients for dimensionless  defects $c_v^{\star} =0.1, 0.2, 0.3, 0.4$. Material parameters have been chosen according to Tab.~SI-1.}
\label{Fig_Comparison}
\end{center}
\begin{center}
\includegraphics[width=0.5\textwidth]{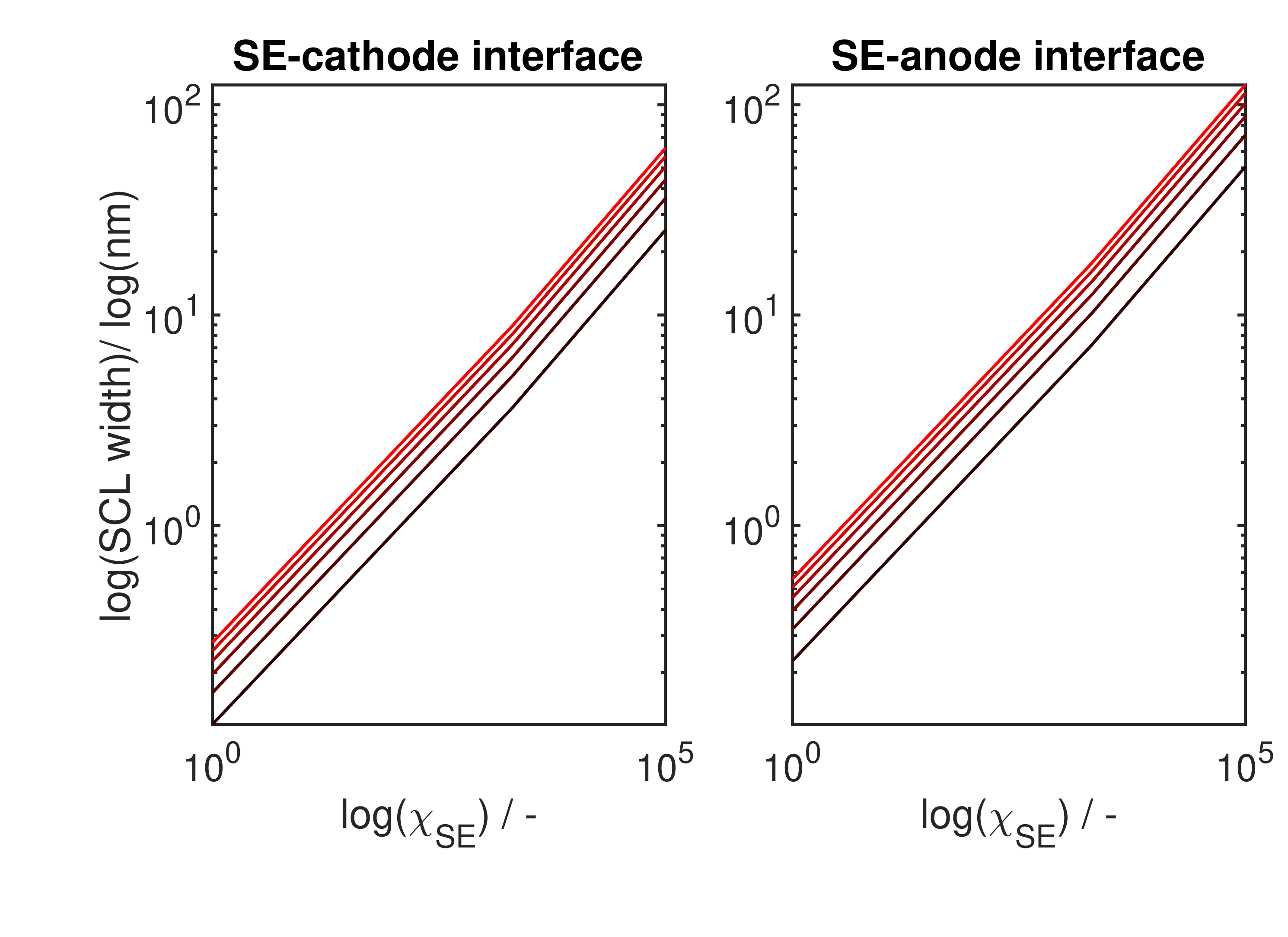}\includegraphics[width=0.5\textwidth]{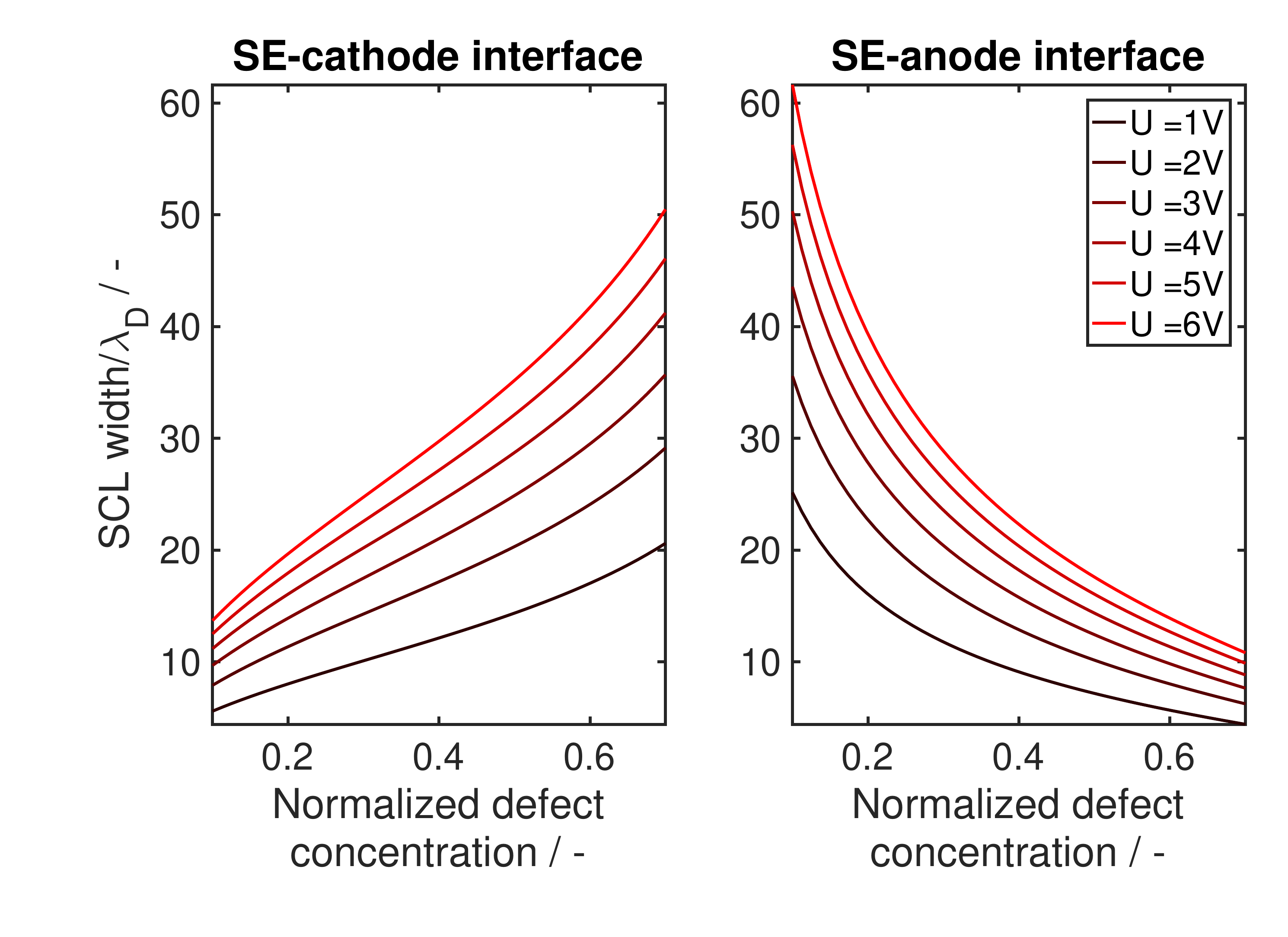}
\caption{SCL widths as a functions of dielectric permittivity (left) and as functions of normalized defect concentration (right): In color gradient $log(l_i)$ versus $log(\chi_{\text{SE}})$ (left) and $l_i/\lambda_D$ versus $c_v/c$ for different applied voltages.}
\label{Fig_SCLWidth2}
\end{center}
\end{figure*}

The absolute height of the bulk plateau $\phi_{\text{bulk}}$ and its value normalized to the applied potential difference $U$ are shown in Fig.~\ref{Fig_BulkPotential} as a function of this potential difference for different amounts of cation sites. For a given material composition, the absolute plateau value (see Fig.~\ref{Fig_BulkPotential} left) increases linearly with respect to the applied potential difference. However, the slope decreases with increasing number of defects (color gradient from dark green to yellow). The plateau value normalized to the potential difference shown in Fig.~\ref{Fig_BulkPotential} right) decreases with increasing defect concentration and tends to a fixed ratio in the limit of large applied potentials. This is due to the observed asymmetry of the SCLs width and the corresponding asymmetric potential drops over the SCLs,  
which can be explained as follows: The more  cation sites are available, the more charge can be stored in accumulation zones with the same thickness. To guarantee global charge neutrality, the corresponding depletion zone spatially expands which results in lower plateau values. Thus, doping of SEs by introducing defects into the lattice structure not only eliminates bottlenecks in the conduction pathways and facilitates transport, but also influences the SCLs.
By changing the amount of  defects the asymmetry can be shifted from the depletion zone to the accumulation zone and vice versa.
If SCLs have an influence on the interfacial resistances, this influence can thus be tuned by the number of  defects between the electrode interfaces.

To further illustrate this point, we vary the number of available cation sites in our simulations from 10\% to 40\%. The simulated potential and concentration profiles in 2400 nm-thick LLTO and LATP layers near the electrode interfaces are shown in Fig.~\ref{Fig_Comparison}. 
Independent of the other material properties the equilibrium bulk potential level decreases with increasing number of cation sites (Fig.~\ref{Fig_Comparison} left). 
This is accompanied with changing SCL potential drops. 
For SEs with 20\% available cation sites the potential drops are non-symmetric. Whereas in the case of 40\% cation sites the potential drops are almost equal. Independent of the amount of cation sites on the negative electrode all sites are filled with cations (see Fig.~\ref{Fig_Comparison} right
). Analogous to the observations regarding the SCL potential drops, the asymmetry of the SCL widths decreases with increasing number of cation sites. 
Finally, we emphasize that the observed asymmetric potential drops at the interfaces is in accordance with the experimental findings of  Y.~Aizawa et al.~\cite{Aizawa_2017}.

\subsubsection{Space-Charge Layer Width}
\label{Sec_LowTemperature}
We observe in our simulations that the width of the SCLs in SEs with spatially homogeneous distributions of cations and cation sites without size effects is much larger than the predicted Debye screening length $\lambda_D$ from Boltzmann-Nernst-Planck theory. In this section we derive an expression for the width from a zero temperature approximation of the gPB equation (\ref{PB2}).

The quasi Fermi electrochemical potential defines a quasi Fermi temperature $\theta_{\Phi} = \Phi/R$, which is in the order of $10^4$~K for effective potential differences from 0.1~V to 6~V and thus very large compared to room temperature. Therefore, to estimate the size of the SCLs, the cation distribution (\ref{Eq_CationDistribution2}) is approximated by its limit $\theta\to 0$,  i.e.,
\begin{subequations}
\begin{alignat}{3}
c_+(\phi)\to & 0\qquad &\text{for }\phi>\Phi, \label{LowTemp1}\\
c_+(\phi)\to &  c_+^{\text{max}},\qquad &\text{for } \phi<\Phi. \label{LowTemp2}
\end{alignat}
\end{subequations}
The cation concentration reaches its maximum for electric potentials $\phi$ up to the quasi Fermi electrochemical potential $\Phi$. Cation sites with higher electrical energy are unoccupied and we can split the SE domain into three electric different regions: the two SCLs and the bulk SE. 
In these regions the gPB equation (\ref{PB2}) reduce to linear gPB equations and  Laplace's equation, respectively
\begin{subequations}
\begin{alignat}{4}
-\epsilon_{\text{SE}}\partial_{xx}  \phi_{\text{C}} &= q_{\text{C}}, \quad & x &\in[0, x_{\text{C}}], \label{LPB11}\\
-\epsilon_{\text{SE}}\partial_{xx}  \phi_{\text{bulk}} &= 0 , \quad &x&\in[x_{\text{C}}, x_{\text{A}}], \label{LPBbilk1}\\
-\epsilon_{\text{SE}}\partial_{xx}  \phi_{\text{A}} &=  q_{\text{A}} \quad & x &\in[x_{\text{A}}, L_{\text{SE}}], \label{LPB21}
\end{alignat}
\end{subequations}
where  $q_{\text{C}} := Fz_0c_0$ and $q_{\text{A}} := z_+Fc_{+}^{\text{max}}+z_0Fc_0$ are the free charge densities in the respective rectangular approximated SCLs. 
These equations can be solved using the potential boundary conditions, global charge conservation, and the requirement of continuity (see Sec.~SI-3.2). The key results are the representations (SI-17) of the locations
\begin{subequations}
\begin{align}
x_{\text{C}}&= \sqrt{2\epsilon_{\text{SE}} U \frac{q_{\text{A}}}{q_{\text{C}}(q_{\text{C}}- q_{\text{A}})}}, \\
x_{\text{A}}& =  L_{\text{SE}} -\sqrt{2\epsilon_{\text{SE}} U\frac{q_{\text{C}}}{ q_{\text{A}}(q_{\text{C}}-q_{\text{A}})}}.\label{locations}
\end{align}
\end{subequations}
These locations give explicit representations of the width of the SCLs $l_{\text{C}}=x_{\text{C}}$ and $l_{\text{A}}=L_{\text{SE}} - x_{\text{A}}= -q_{\text{C}}/q_{\text{A}} l_{\text{C}}$ as material and operation mode dependent functions. 
\begin{itemize}
\item
For fixed voltage $U$, the SCL widths are different from the Debye length of order 
\begin{subequations}
\begin{align}
l_{\text{C}}&\sim{\cal O}\left(\sqrt{\epsilon_{\text{SE}} \frac{c^{\text{max}}_+-c_0}{c_0c^{\text{max}}_+}}\right), \label{ESCLwidth1}\\
l_{\text{A}}&\sim{\cal O}\left(\sqrt{\epsilon_{\text{SE}} \frac{c_0}{c_+^{\text{max}}(c_+^{\text{max}}-c_0))}}\right).\label{ESCLwidth2}
\end{align}
\end{subequations}
They scale as $\lambda_D$ with the root of the dielectric property but also depend on the composition and are different for anode and cathode.   
\item
Depending on the chosen dielectric susceptibility $\epsilon_{\text{SE}}$ the predicted space charge width are either a few nanometer or up to 200~nm, if zero frequency limits of the dielectric constants are the relevant effective dielectric constants for the calculation of space charge layers. As discussed in Sec.~SI-2.1, the exact choice of the measured dielectric properties for the parametrization of continuum models is subject of current debates and research.
Fig.~\ref{Fig_SCLWidth2} (right) illustrates the qualitative dependence of the SCLs width for dielectric permittivities $\chi_{\text{SE}}$ from $10^{1}$ to $10^{5}$ in a log-log plot.  
\item
The width of the depletion and accumulation zones scale differently with composition. Regardless of the dielectric susceptibility selected, this leads on the one hand to the asymmetric behavior of the SCLs and to significantly larger SCLs than in other electrochemical systems. 
Fig.~\ref{Fig_SCLWidth2} (right) shows the dimensionless factors by which the spatial extension of the SCLs deviate from the Debye length as a function of normalized defect concentrations.  
These findings are consistent with our previous simulation results. For a defect concentration of $c_v/c =0.2$ and a potential difference of 2 V, the absolute value of the width is 10 times as large as the Debye length. This effect on the depletion zone increases with increasing defect concentration. The behavior of the accumulation zones  (Fig.~\ref{Fig_SCLWidth2} right) is opposite but not symmetrical. The dimensionless factors for both the depletion zone  
and the accumulation zone 
are depicted for a variation of effective potential differences from 1~V to 6~V. The higher the potential difference, the greater the deviation from the behavior predicted for monovalent dilute liquid electrolytes.
\item
For the sake of completeness, we mention that the low temperature approximation also gives an approximation (SI-18) of the bulk potential.
Combining this with the bulk chemical potential $\hat{\mu}_{\text{bulk}} =\hat{\mu}^{\circ} +R\theta\ln(c_0/c_v)$ we obtain an approximation for the quasi Fermi potential
\begin{align}
\Phi &=  \frac{R\theta}{z_+F}\ln\left(\frac{c_0}{c_v}\right)+ U \frac{q_{\text{C}}}{q_{\text{C}}-q_{\text{A}}}.
\end{align}
Particularly, this expression shows explicitly why the incline of the bulk potential observed in Fig.~\ref{Fig_BulkPotential} approaches a constant value. 
\end{itemize}

\subsubsection{Differential Space-Charge Layer Capacity}
Finally, let us turn to space charge storage. The response of the charge stored in the SCLs to changes in the potential drop across the SCL is characterized by the SCL differential capacities 
\begin{align}
C_{i}= -\frac{d Q_{i}}{d\Delta \phi_{i}},\qquad  i \in\{\text{A, C}\},\label{SCLCapacity1}
\end{align}
where $\Delta \phi_{i} =\phi_{\text{bulk}}- \phi_{\text{bnd}}$ denotes the SCL potential drop between the bulk SE and the interface and the stored net charge $Q_{i}$ is given by the integrated free charge density of the SCL.

Here, we employ an approach by Bazant et al. \cite{Bazant_2009} that utilizes that  
the free charge density is not only a function of the electric field, but also a function of the pressure via the force balance to calculate the SCL differential capacity. This is based on the reformulation of the differential capacitance  (\ref{SCLCapacity1}) with chain rule
\begin{align}
C_{i}= -\frac{d Q_{i}}{d \Delta p_{i}} \frac{d \Delta p_{i}}{d\Delta \phi_{i}},\qquad  i \in\{\text{A, C}\},\label{SCLCapacity2}
\end{align}
where $\Delta p_{i}= p_{\text{bulk}}-p_{\text{bnd}}$ denotes the SCL pressure drops. Using the analytical expressions for the electric field given in the supporting information we obtain the space charge
as function of this pressure difference. 
We find a non-linear analytical expression for the differential capacitance, which is a result of the material model stated in Sec.~\ref{Subsec_SEModel}.  For details see Sec.~SI-3.3.

\begin{figure}
\begin{center}
\includegraphics[width=0.5\textwidth]{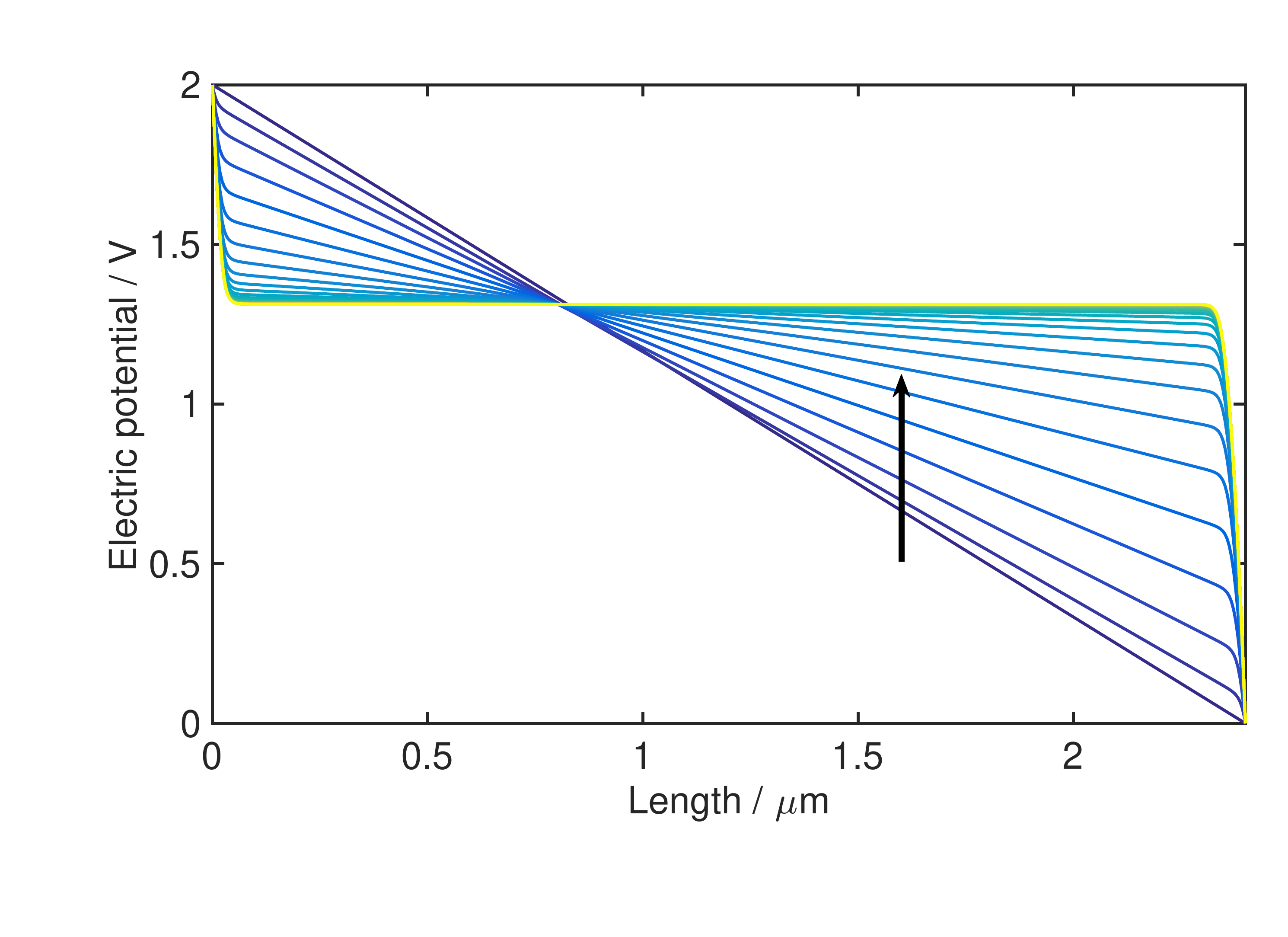}
\includegraphics[width=0.5\textwidth]{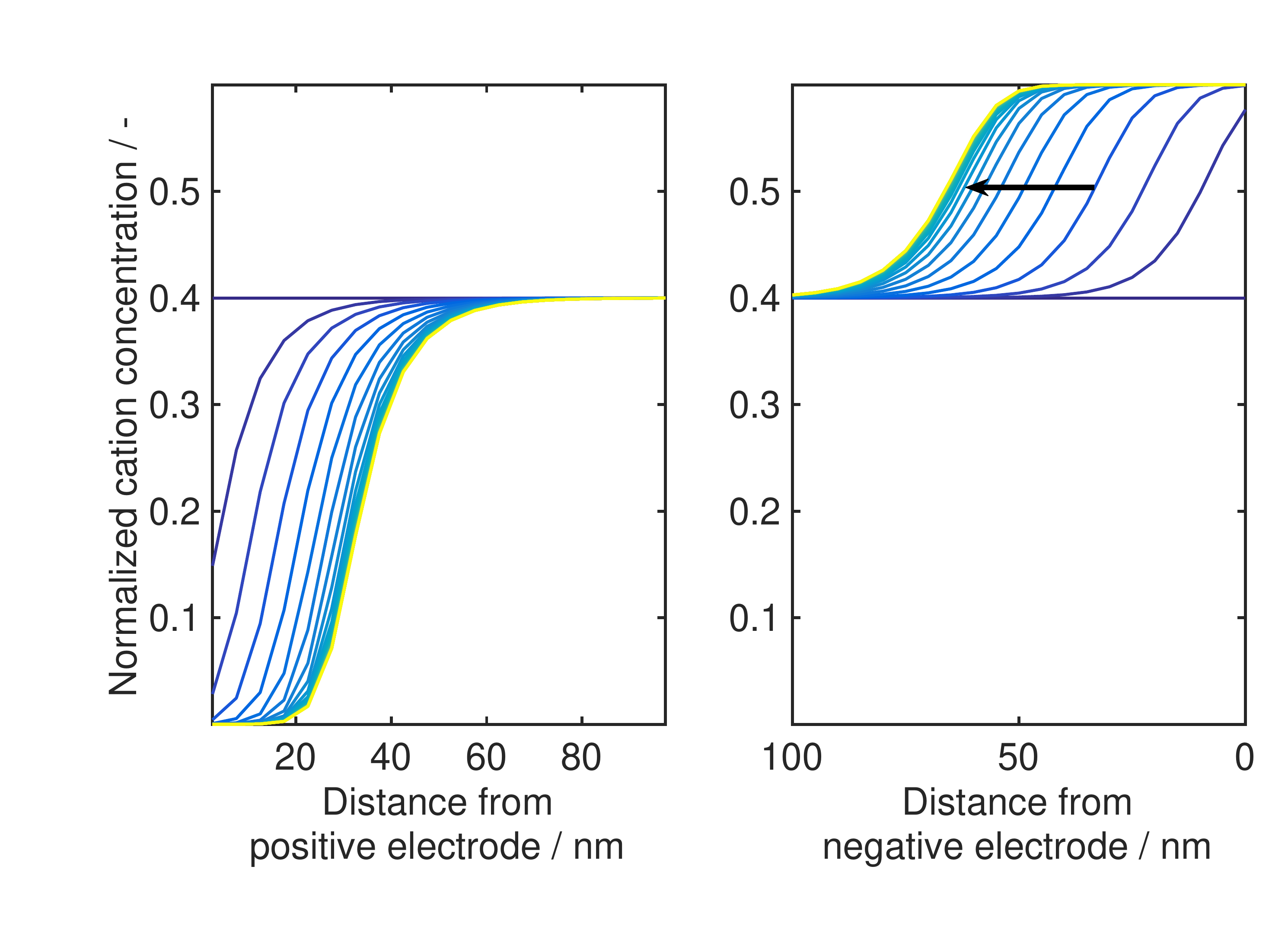}
\includegraphics[width=0.5\textwidth]{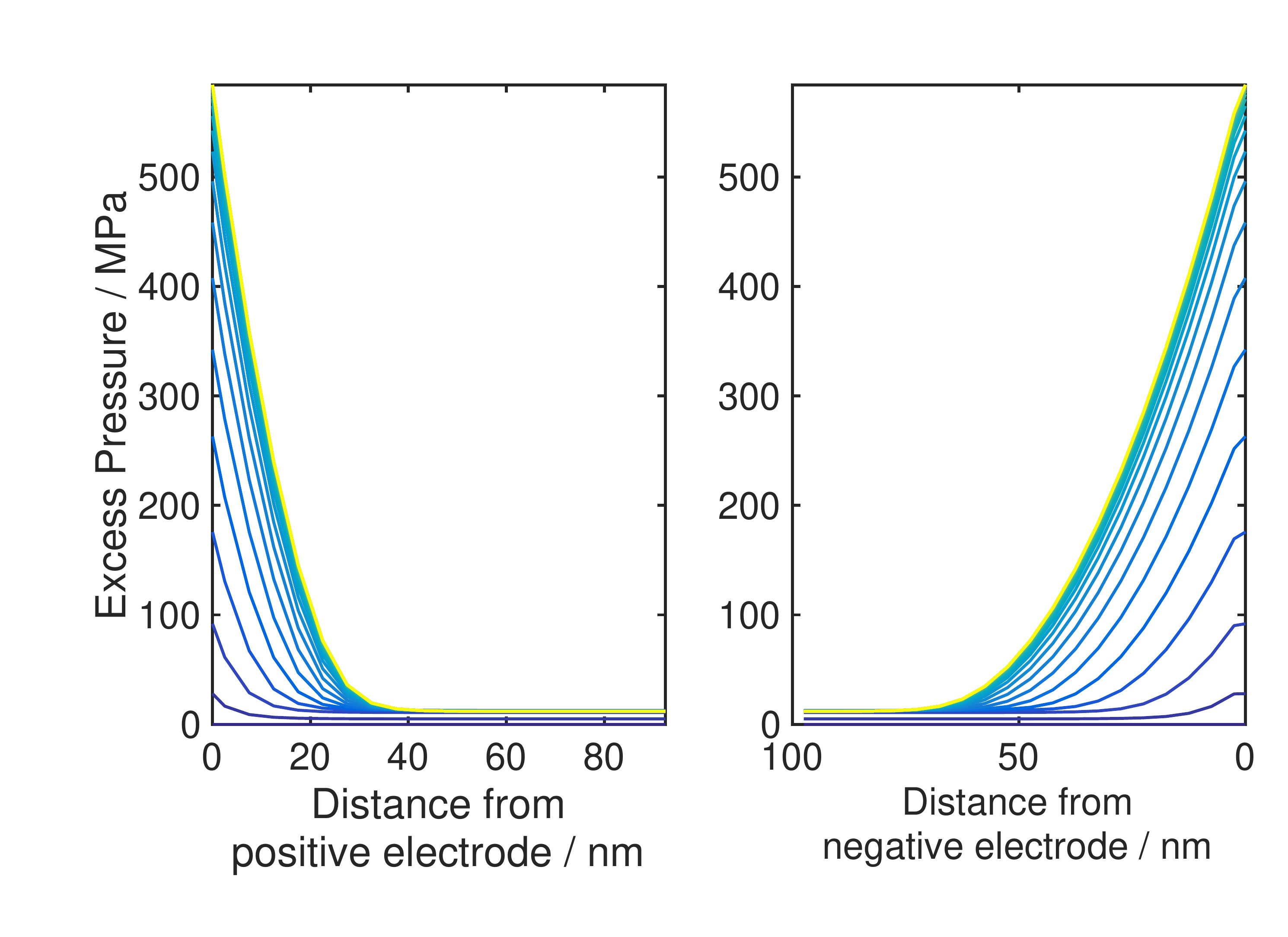}
\caption{Predicted time evolution of electric potential (top) and cation concentration near the interfaces (bottom) in 2400 nm-thick LLTO to a sudden applied voltage step of 2 V at selected times. 
Over time both state variables evolve until in equilibrium is reached (yellow lines). Material parameters have been chosen according to Tab.~SI-1.}
\label{Fig_PotentialEvolution}
\end{center}
\end{figure}

\begin{figure*}
\includegraphics[width=0.5\textwidth]{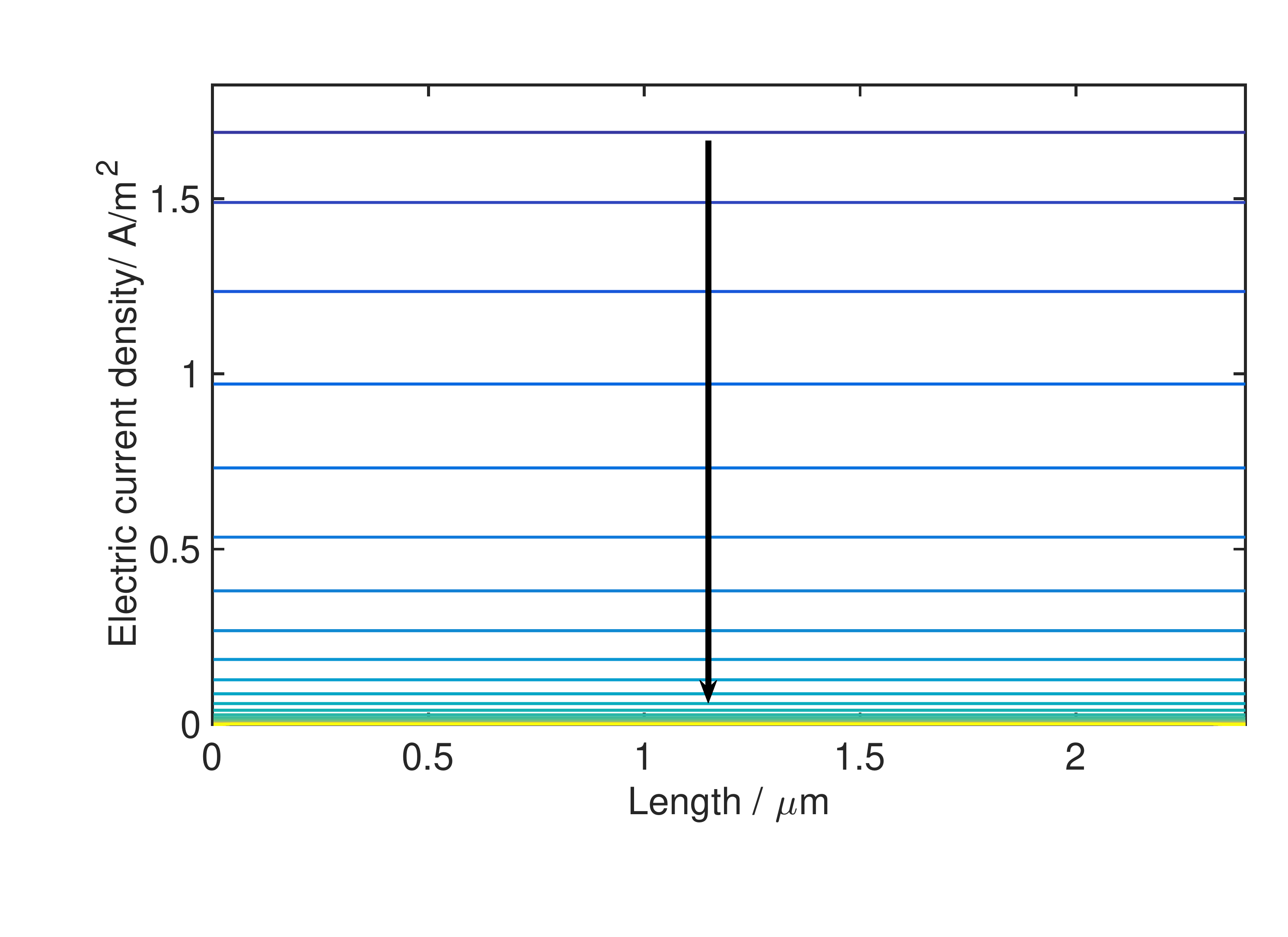}\includegraphics[width=0.5\textwidth]{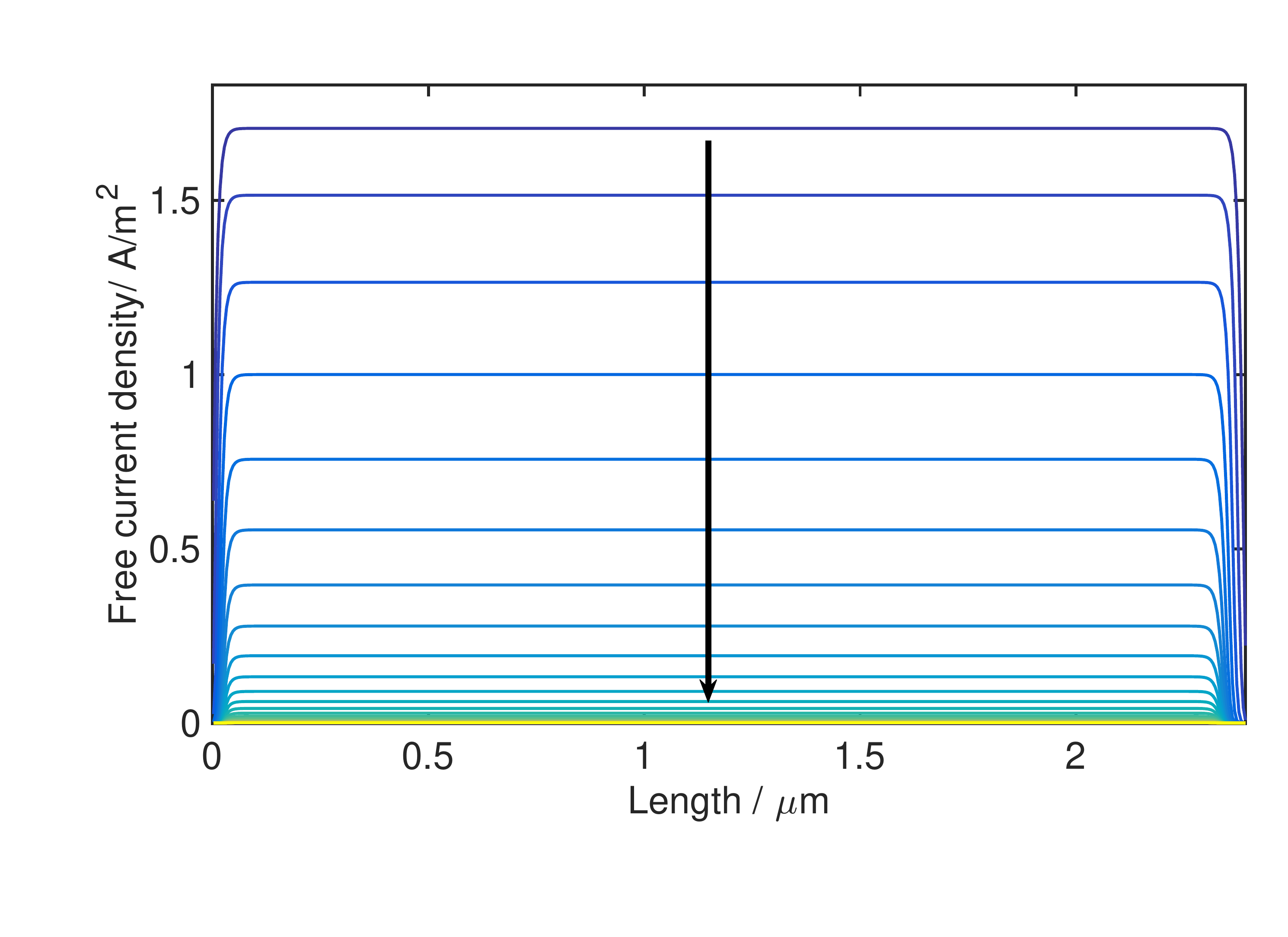}
\includegraphics[width=0.5\textwidth]{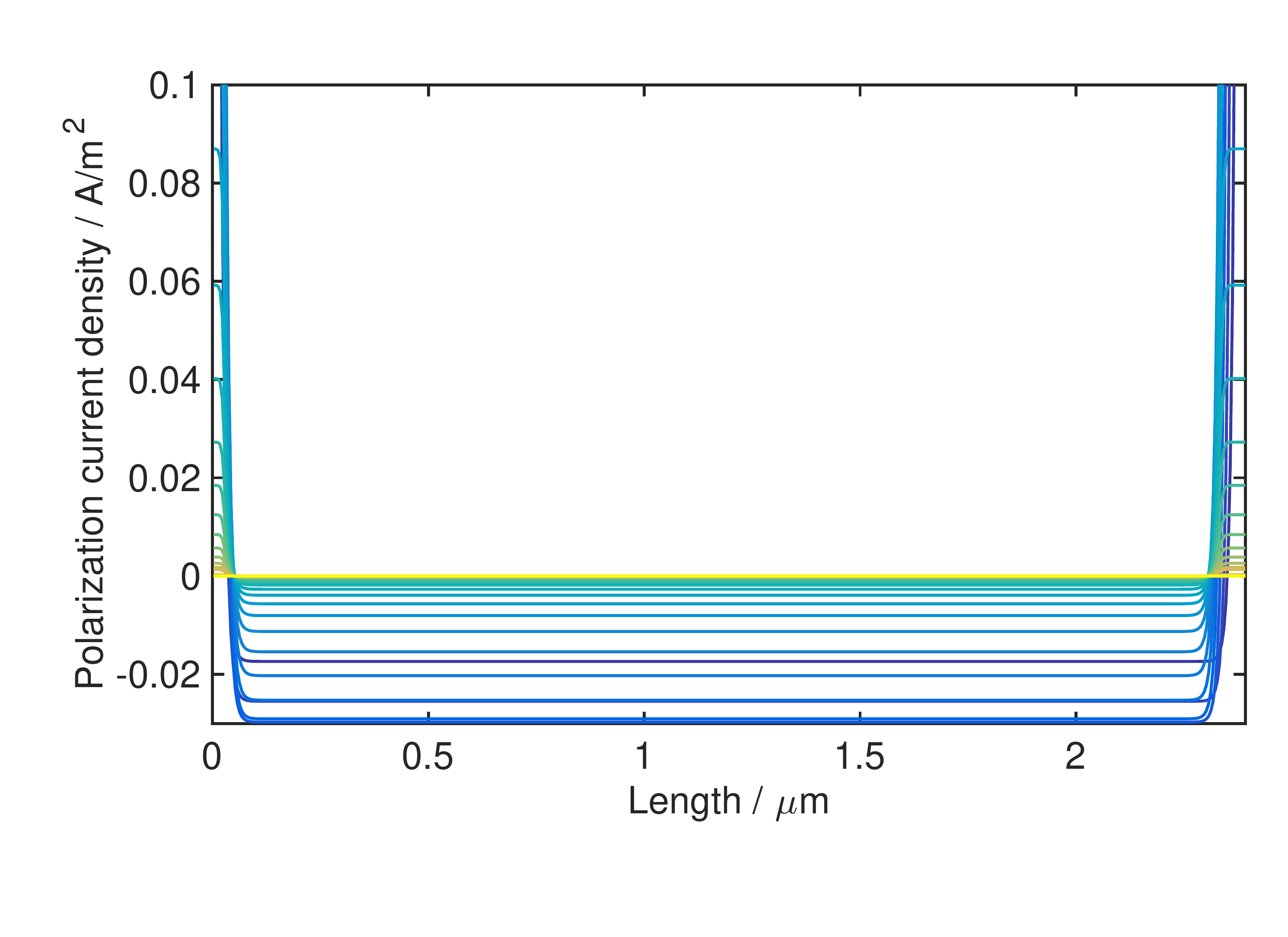}\includegraphics[width=0.5\textwidth]{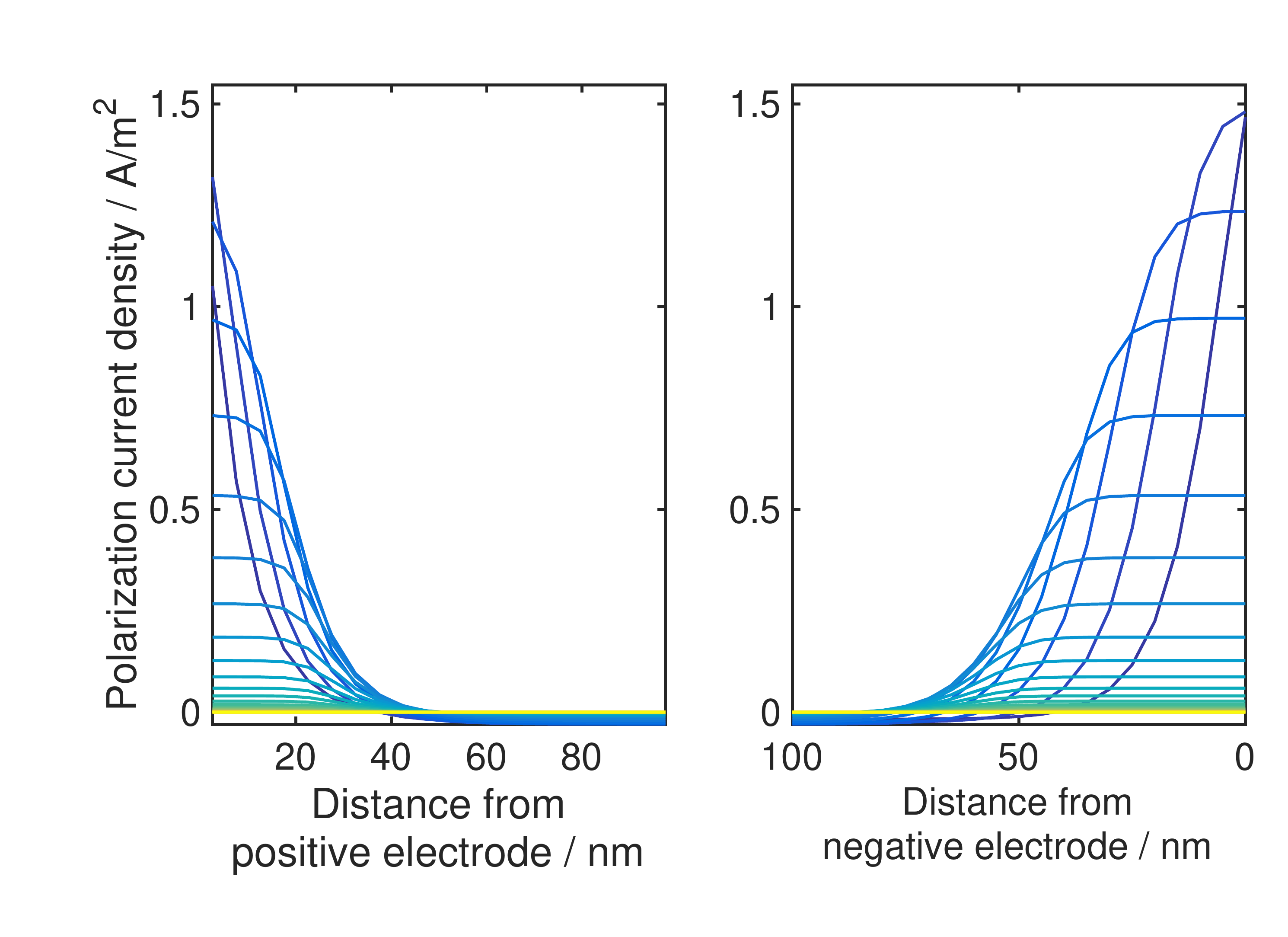}
\caption{Contributions to the electric current density response in 2400 nm-thick LLTO induced by a sudden applied voltage step of 2~V  at selected times. The electric current response (top left) is linear and decreases over time until equilibrium is reached (blue to yellow). The main part of the electric current density in the bulk SE is caused by the free current density (top right), while the electric current density in the SCLs is mainly generated by the polarization current density (bottom). Material parameters have been chosen according to Tab.~SI-1.}
\label{Fig_CurrentContributions}
\end{figure*}

\subsection{Relaxation Phenomena}
\label{Subsec_Dynamik}
As a case study, we consider a 2400 nm-thick LLTO, which is a prominent example of a highly polarizable~\cite{Bucheli_2014} ceramic SE with high bulk and total lithium-ion conductivity\cite{Kwon_2017}. We illustrate the physics of polarization induced processes perpendicular to some planar ion-blocking electrodes by 1D numerical results.

\subsubsection{Time Evolution of State Variables}
Fig.~\ref{Fig_PotentialEvolution} (top) shows the time evolution of the electric potential distribution in LLTO.
The solutions are plotted at different times. Starting at $t=0$, when the SE is brought in contact with the electrodes no electric field is present (dark blue line). 
Then, due to the polarization, a linear a constant electric field is build up.
Fig.~\ref{Fig_PotentialEvolution} (bottom) illustrates the cation concentration dynamics near the electrode interfaces. Initially at $t=0$, the cations are distributed homogeneously throughout the LLTO lattice (dark blue lines). Both the cation concentration and the potential evolve in time until equilibrium is reached (yellow lines).
In equilibrium, the cations have been diffused and forced away by the electric field resulting in cation depletion regions next to the positive electrode (yellow line in Fig.~\ref{Fig_PotentialEvolution} left bottom) and cation accumulation regions next to the negative electrode (yellow line in Fig.~\ref{Fig_PotentialEvolution} right bottom).  In contrast to liquid electrolytes\cite{Dreyer_2013}, SEs show SCLs of very few cations on the positive electrode side and very few vacancies on the negative electrode side. Like in liquid electrolytes, the pressure correction entering the chemical potentials along with the incompressibility constraint prevent overestimating the cation concentration at the SE-electrode interfaces. In the bulk material concentration of cations and anions are identical leading to local charge neutrality. Due to the charge redistribution the potential relaxes until the electric field appears only in the boundary layers. Steady state is characterized by charge neutrality and a potential plateau in the bulk material, high potential drops in the boundary layers. Because of the chosen high (low frequency) dielectric susceptibility we observe large SCLs in the range of 50-100 nm.

\subsubsection{Time Evolution of Elastic Pressure}
In our simulations the total external stress acting on the SE-electrode interfaces is set to ambient pressure and we assume, that no charge is absorbed at the interfaces. Both assumption can easily be changed in the simulation. The continuity of the stress tensor at the SE-electrode interfaces then allows us to calculate the evolution of the mechanical pressure in the solid electrolyte pressure (see Fig.~\ref{Fig_PotentialEvolution}), which is acting on the anion lattice due to the build up of electric fields and corresponding Maxwell stress in the SCLs. 
In equilibrium, the strong electric fields in the two boundary layers are counterbalanced by the elastic pressure in the range of MPa to obtain a constant total stress. 
\subsubsection{Electric Current Response}
Fig.~\ref{Fig_CurrentContributions} (top left) shows the electric current density evolution at selected times. The total electric current density response to the applied voltage is constant through the SE and decreases over time until equilibrium is established (yellow line). As discussed in Sec.~\ref{Subsec_SEModel}, this current density contains two contributions: the free current density and the polarization current density. We realize from Fig.~\ref{Fig_CurrentContributions} (top right) that the main part of the total electric current density in the bulk SE comes from the free current density, whereas in the SCLs, cation motion does not contribute significantly to the electric current density. In the SCLs the major current density contribution is generated by the polarization response
(see Fig.~\ref{Fig_CurrentContributions} bottom right). The essential parts of both current densities decrease over time until in equilibrium the total electric current density vanishes. Notably, the contribution of the bulk polarization current increases slightly over time (see Fig.~\ref{Fig_CurrentContributions} bottom left). This counteracts the decrease of the bulk free current density and delays the overall current reduction. 

\section{Summary and Conclusions}\label{Sec_Conclusion}
The width of the SCLs and their influence on the high interfacial resistances in ASSB cells are currently controversially discussed.  
Many modeling approaches for the description of processes in SEs and the near interface regions are based on simplified phenomenological descriptions and neglect the interaction between elastic pressure and Maxwell stresses. In this article, we present a free energy based continuum model for SE dynamics. The mathematical model is able to predict coupled dynamic effects in both the bulk and the near interface regions without assumptions on the structure of interfaces and SCLs. Our main \emph{results} derived from this model are:
 \begin{enumerate}
\item Statics: Unlike in liquid electrolytes, the SCLs are asymmetric and their width does not only scale with the dielectric properties of the SE, but also with the electrolyte composition and in particular the concentration of cation sites and defects.
We find, in agreement with the experimental observations~\cite{Yamamoto2010, Hirayama_2017, Aizawa_2017, Nomura_2019}, that regardless of the order of magnitude of the dielectric susceptibilities, the SCLs in SEs are wider than in liquid electrolytes and exceed the classical Debye screening length by at least one order of magnitude. The width of the SCLs may be defined by the quasi Fermi electrochemical potential established in equilibrium. This indicates how the width of the SCLs can be influenced by bulk properties. Furthermore, our analytical results for the bulk potential allow us to calculate the asymmetric interfacial potential drops. This result is again qualitatively in line with the experimental works~\cite{Yamamoto2010, Hirayama_2017, Aizawa_2017, Nomura_2019}.

\item Dynamics: The process of SCL formation is accompanied with various boundary layer phenomena, such as pronounced electric fields and high elastic pressures. Due to the coupling of mechanical and Maxwell stress, we reveal high forces acting on the solid-solid interfaces. In contrast to bulk type models and many phenomenological SCL resolved approaches, we describe the different contributions to the electric current. The major part of electric currents in the SCLs in non-stationary conditions is caused by the polarization current.
\end{enumerate}
As in all continuum models for phenomena in dielectrics, the \emph{quantitative results} predicted by our model, depend strongly on the magnitude of dielectric susceptibilities. If our theoretical conjecture is correct that at finite electrical currents, the relevant dielectric constant for the calculation of SCLs in electrolytes with grain boundaries is closer to the low frequency limit of the dielectric response function in crystalline SEs,
LLTO perovskites will develop over time large asymmetric SCL of 50-100 nm width at both electrode interfaces. In contrast, for amorphous glasses such as LiPON, where the high frequency response is relevant, the predicted SCLs width are in the order of nanometers. It remains to be shown whether this conjecture can be confirmed by experiment.

Finally, under the assumption of mechanical equilibrium, we trace the model derived by Braun et al.~\cite{Braun_2015} back to the model introduced by Kornychev and Vorotyntsev~\cite{Kornychev_1981}, which does not consider mechanical effects. In particular, we show that this does not apply to inhomogeneous conduction pathways. Thus, this paper generalizes the established models and gives new insights in predictions.

\section*{Conflicts of interest}
There are no conflicts to declare.

\section*{Acknowledgements}

Financial support from the Federal Ministry of Education and Research (BMBF) within the FELIZIA project (03XP0026F) and the FestBatt project (03XP0174C) is gratefully acknowledged. This work contributes to the research performed at CELEST (Center for Electrochemical Energy Storage Ulm-Karlsruhe).

\begin{table}[h]
\small
  \caption{\ Abbreviations.}\label{Tab_Abbreviations}
  
  \begin{tabular*}{0.48\textwidth}{@{\extracolsep{\fill}}ll}
  \hline
  \textbf{Abbreviation} & \textbf{Phrase}\\
    \hline
    &\\
ASSB & All-solid-state battery\\
LATP & Li$_{1+x}$Al$_{x}$Ti$_{2-x}$(PO$_4$)$_3$ (at $x=0.3$)\\
LiPON & LiPO$_{4-x}$N$_x$ (at $x=0$)\\
LLTO & Li$_{0.5-3x}$La$_{0.5+x}$TiO$_3$ (at $x=0$)\\
NaSICON & (Na$^+$) super ion conductor\\
PF & Poisson-Fermi\\
PNP & Poisson-Nernst Planck\\
SCL & Space-charge layer\\
SE & Solid electrolyte\\
&\\
\hline
\end{tabular*}
\end{table}

\begin{table}[h]
\small

\caption{\ Nomenclature and description of physical constants.}\label{Tab_Constants}  
  \begin{tabular*}{0.48\textwidth}{@{\extracolsep{\fill}}lll}
    \hline
  \textbf{Symbol}  & \textbf{Value Unit} & \textbf{Description}\\
    \hline
    &&\\  
$\epsilon_0$  &8.85$\cdot 10^{-12}$ F/m& Dielectric permittivity of vacuum\\
$F$ & 9.65$\cdot 10^{4}$ As/mol & Faraday constant\\
$R$ &8.314 J/(mol K)& Universal gas constant\\
$p_0$ & 101 325 Pa & Standard atmospheric pressure\\
$\theta$ & 298 K&  Room temperature\\
&&\\
\hline\\
\end{tabular*}
\end{table}
\begin{table}
\small

\caption{\ Nomenclature and description of quantities.}\label{Tab_Symbols}
\begin{tabular*}{0.48\textwidth}{@{\extracolsep{\fill}}lll}
    \hline
   \textbf{Symbol} &  \textbf{Description} &\textbf{Unit} \\
    \hline
    &\\
 $\beta$ & Duality factor &-\\
$b_+$ & Electrical cation mobility & mol/(V m s)\\
$c_{\alpha}$& Molar densities of species & mol/m$^3$\\
$c_1$ &Mobile molar density& mol/m$^3$\\
$c$ &Summary molar density& mol/m$^3$\\
$c^{\text{max}}_+$ &Maximum cation concentration& mol/m$^3$\\
$D_+$ &Chemical diffusion coefficient &m$^2$/s\\
$D_+^t$ &Tracer diffusion coefficient &m$^2$/s\\
$D_p$ &Baro-diffusion coefficient & mol s/kg\\
$D_{ih}$ &Inhomogeneity parameter &m$^2$/s\\
   $\textbf{D}$ &Dielectric displacement field &As/m$^2$\\
    $\textbf{E}$ &Electric field &V/m\\
    $\epsilon_{\text{SE}}$ & Dielectric susceptibility &A$^2$s$^4$/(kg m$^3$)\\
    $f_{\alpha\beta}$ & Thermodynamic correction factors &-\\
    $\gamma_{\alpha}$ &Activity coefficients of species &m$^3$/mol\\
    $\textbf{j}^{\#}$ & Electric current density vector &A/m$^2$\\
$\textbf{j}_{F}^{\#}$ & Free current density vector &A/m$^2$\\
$\textbf{j}_{P}$ & Polarization current density vector &A/m$^2$\\
$K_{\text{SE}}$ & Bulk modulus &kg/(m s$^2$)\\
$L_{\text{SE}}$  &SE length &m\\
$L_{\text{diff}}$ & Diffusion lenght&m\\
$\lambda_D$ & Debye length&m\\
$l_i$ &SCL width&m\\
${\cal L}_{\alpha\beta}$ &Onsager coefficients &mol$^2$/(J m s)\\
$M_{\alpha}$ &Molar masses of species &kg/mol\\
$m_+$ & Ratio of molar masses &-\\
$\mu_{\alpha}$ &Chemical potentials of species&J/mol\\
$\tilde{\mu}$ & Effective chemical potential &J/mol\\
$N$ &Number of species &-\\
$\textbf{N}_{\alpha}^{\#}$ & Total molar fluxes of species (lab frame)& mol/(m$^2$s)\\ 
$\textbf{N}_{\alpha}$ &Molar diffusion fluxes of species & mol/(m$^2$s)\\
$\nu_{\alpha}$ &Partial molar volumes of species& m$^3$/mol\\
$p$ &Elastic pressure& Pa=kg/(m s$^2$)\\
$\textbf{P}$ &  Polarization vector & V/m\\
$\phi$ & Electro-static potential &V\\
$\Phi$ & Quasi Fermi electrochemical potential &J/mol\\
$\tilde{\varphi}$ &Effective electrochemical potential& J/mol\\
$q_F$ & Free charge density &As/m$^3$\\ 
$q$ & Total electric charge density & As/m$^3$\\
$\rho_{\alpha}$  & Mass densities of species &kg/m$^3$\\
$\rho$ & Mass density &kg/m$^3$\\
$\rho\psi$ &Free energy density&J/m$^3$\\
$\sigma$ &Conductivity & S/m =A/(Vm)\\
$t $ & Time &s\\
${\cal T}$ & Stress tensor &N/m$^2$\\
$\theta_{\Phi}$ & Fermi temperature&K\\
$U$ & Applied voltage &V\\
$\textbf{v}_{\alpha}$  & Velocity of species &m/s\\
$\textbf{v}$  & Center of mass velocity &m/s\\
$x_i$ & Positions &m\\
$\textbf{X}_{\alpha}$ &Driving force of species&J/(mol m)\\
$\overline{\textbf{X}}$ &Effective driving force & J/(mol m)\\
$\chi_{\text{SE}}$ &Dielectric susceptibility&-\\
$z_{\alpha}$  &Charge number of species&-\\
$\overline{z}$ &Effective charge number&-\\
$\zeta$ &Entropy production& W/m$^3$=kg/(m s$^3$)\\
&\\
\hline\\
\end{tabular*}
Superscripts `$\circ$' indicate quantities with respect to the reference configuration. Superscripts `$R$' and `$\star$' mark reference and dimensionless quantities in the context of dimensional analysis.  Quantities associated with the lattice-fixed frame are marked with a $\#$ and quantities in mechanical equilibrium are indicated by a hat. 
Sup- and superscripts `$\text{bulk}$' label bulk quantities, whereas subscripts `A' and `C' denote quantities in the anodic and cathodic SCLs. Supscripts `L' and `R' indicate boundary values.
\end{table}

\bibliography{SEArticle} 
\bibliographystyle{rsc} 

\end{document}